\documentclass[11pt,english]{article}
\usepackage[T1]{fontenc}
\usepackage{amsmath, amsthm, amsfonts}
\usepackage[hidelinks]{hyperref}
\usepackage{cleveref}
\Crefformat{appendix}{\S#2#1#3}

\usepackage{framed}
\usepackage{todonotes}
\makeatletter
\define@key{todonotes}{AC}[]{%
    \setkeys{todonotes}{author=AGC,color=blue!25}}%
\makeatother
\usepackage{paralist}
\usepackage{url}

\newtheorem{assumption}{Assumption}
\newtheorem{definition}{Definition}[section]
\newtheorem{example}{Example}[section]
\usepackage[margin=.8in,footskip=0.25in]{geometry}

\definecolor{cb-black}      {RGB}{  0,   0,   0}
\definecolor{cb-blue-green} {RGB}{  0,   73, 73}
\definecolor{causal}        {RGB}{  0, 146, 146}
\definecolor{cb-pink}       {RGB}{255, 109, 182}
\definecolor{cb-salmon-pink}{RGB}{255, 182, 119}
\definecolor{cb-purple}     {RGB}{ 73,   0, 146}
\definecolor{cb-blue}       {RGB}{ 0, 109, 219}
\definecolor{cb-lilac}      {RGB}{182, 109, 255}
\definecolor{adjusted}      {RGB}{109, 182, 255}
\definecolor{cb-blue-light} {RGB}{182, 219, 255}
\definecolor{association}   {RGB}{146,   0,   0}
\definecolor{cb-brown}      {RGB}{146,  73,   0}
\definecolor{cb-clay}       {RGB}{219, 209,   0}
\definecolor{cb-green-lime} {RGB}{ 36, 255,  36}
\definecolor{cb-yellow}     {RGB}{255, 255, 109}

\usepackage{tikz}
\usetikzlibrary{arrows, positioning}
\tikzset{
    vertex/.style={circle,draw,minimum size=1.8em, thick},
    edge/.style={->,> = latex', thick}
}
\usepackage{nicefrac}
\usepackage{booktabs}
\usepackage{multirow}
\usepackage{listings}
\usepackage{courier}
\DeclareMathOperator{\E}{\mathbb{E}}
\usepackage{dsfont}
\usepackage{soul}

\newcommand*\circled[1]{\tikz[baseline=(char.base)]{
            \node[shape=circle,draw,inner sep=1pt, thick] (char) {$\mathtt{#1}$};}}

\newcommand*{\expe}{\mathds{E}}

\author{
  Andrew G. Clark\\
  \small\texttt{agclark2@sheffield.ac.uk}
  \and
  Michael Foster\\
  \small\texttt{m.foster@sheffield.ac.uk}
  \and
  Benedikt Prifling\\
  \small\texttt{benedikt.prifling@uni-ulm.de}
  \and
  Neil Walkinshaw\\
  \small\texttt{n.walkinshaw@sheffield.ac.uk}
  \and
  Robert M. Hierons\\
  \small\texttt{r.hierons@sheffield.ac.uk}
  \and
  Volker Schmidt\\
  \small\texttt{volker.schmidt@uni-ulm.de}
  \and
  Robert D. Turner\\
  \small\texttt{r.d.turner@sheffield.ac.uk}
}

\title{Testing Causality in Scientific Modelling Software}

\date{}

\begin{document}
\maketitle

\begin{abstract}
From simulating galaxy formation to viral transmission in a pandemic, scientific models play a pivotal role in developing scientific theories and supporting government policy decisions that affect us all. Given these critical applications, a poor modelling assumption or bug could have far-reaching consequences. However, scientific models possess several properties that make them notoriously difficult to test, including a complex input space, long execution times, and non-determinism, rendering existing testing techniques impractical. In fields such as epidemiology, where researchers seek answers to challenging causal questions, a statistical methodology known as Causal Inference has addressed similar problems, enabling the inference of causal conclusions from noisy, biased, and sparse data instead of costly experiments. This paper introduces the Causal Testing Framework: a framework that uses Causal Inference techniques to establish causal effects from existing data, enabling users to conduct software testing activities concerning the effect of a change, such as Metamorphic Testing, \emph{a posteriori}. We present three case studies covering real-world scientific models, demonstrating how the Causal Testing Framework can infer metamorphic test outcomes from reused, confounded test data to provide an efficient solution for testing scientific modelling software.
\end{abstract}



\def\arraystretch{0.8}

\section{Introduction}\label{sec:introduction}

The use of scientific modelling software to model, simulate, and understand complex phenomena has become commonplace. Such systems have played a pivotal role in improving our scientific understanding across a wide range of phenomena and disciplines, and are increasingly used outside of academia. Governments, for example, make extensive use of scientific modelling software to simulate and evaluate various policies and interventions \cite{oldfield2021analytical}. Perhaps most notably, this has included the use of epidemiological models to predict the impact of a number of COVID-19 mitigation measures \cite{thompson2020epidemiological, kerr2021covasim}.

Testing such models is particularly challenging \cite{kanewala2014testing}. They typically have vast input spaces comprising hundreds of parameters, as well as complex output spaces. Executing large numbers of tests is often impossible, because each execution can require a significant amount of time and resource to execute. Compounding this issue further, scientific models are often non-deterministic, meaning developers must run each test case multiple times and observe the distribution of outputs. Furthermore, these systems are often developed by scientists with a limited amount of training as software engineers  \cite{kelly2008challenge}.

Collectively, these issues make it difficult (and sometimes impossible) to determine whether the output of a test case or modelling scenario is correct or not. This is referred to as the test oracle problem \cite{barr2015}. Instead, to determine whether a software system is fit for purpose, a tester generally corroborates evidence to investigate smaller, more specific relationships between inputs and outputs. By making changes to particular input parameters and observing changes to particular output variables, there is an implicit assumption that the input parameters in question somehow influence the computation (i.e. have a `causal' effect) of the outputs.




In this paper we are specifically concerned with this intrinsic challenge: How can we test the (implicitly causal) input-output relationships in a system with a vast and complex input space, which may be non-deterministic and suffer from the test oracle problem, without the ability to resort to large numbers of test executions?

The challenge of analysing causal relationships in limited, noisy data instead of running costly experiments is well-established in the statistical context. In areas such as epidemiology, a powerful statistical methodology known as causal inference (CI) has been employed to answer causal questions that cannot be answered experimentally due to ethical concerns, such as \emph{Does smoking cause lung cancer?} \cite{cornfield1959smoking}. By incorporating domain knowledge about known causal relationships between variables (or absence thereof), CI can produce \emph{estimands} that isolate the causal quantity of interest. That is, `recipes' for analysing data in a causally-valid way. Conventional statistical methods can then be employed to quantify the presence (or absence) of specific causal relationships, correcting for bias in the data, without the need for experimental procedures.

This paper is motivated by the observation that CI and software testing share a common goal in many cases: to establish precise and salient causal relationships. Moreover, by viewing software testing through a causal lens, we can leverage well-established CI techniques that conceptually address several testing challenges presented by scientific models for causality-driven testing activities, such as metamorphic testing.

To this end, we introduce a testing framework that incorporates an explicit model of causality into the testing process, facilitating the direct application of CI techniques to software testing problems, such as metamorphic testing. To achieve this, we take a model-based testing (MBT) perspective \cite{moore56}, in which testing is based on a model of the expected behaviour of the system-under-test that typically either describes the allowed sequences of events or gives a formal relation between the inputs and outputs~\cite{HieronsBBCDDGHKKLSVWZ09,UttingPL12}. Traditionally, MBT has focused on models expressed using state-based languages, such as finite state machines \cite{lee96} and labelled transition systems \cite{Tretmans08}, or models that define the allowed input-output relationships using languages, such as Z \cite{hierons97b} and VDM \cite{dick93}. However, given the focus on causality in this paper, we require a model that specifies the expected \emph{causal relationships} between system inputs and outputs.
Here, we assume that such causal information is represented by a causal directed acyclic graph (DAG) \cite{pearl1995, hernan2020causal}.


Our decision to incorporate causal DAGs into the testing process is motivated by two main factors. First, testing can be viewed as a causal activity in which the tester checks whether expected causal relationships hold; in order to automate this process, we require the expected causal relationships to be expressed. Second, the causal DAG is a lightweight and intuitive model that is widely used by domain experts in areas such as epidemiology and social sciences to make causal assumptions actionable and transparent \cite{greenland1999causal, tennant2021use}. 


In this paper, we make three contributions. First, we introduce a conceptual framework that approaches software testing activities, such as metamorphic testing, as CI problems, and clarifies the components necessary to leverage state-of-the-art CI techniques. While previous work \cite{bareinboim2016} has shown that CI is, generally speaking, a universally applicable technique, we believe we are the first to apply it to the software testing field in this way. Second, we provide a reference implementation of the framework that can form the basis for future CI-driven tools for testing scientific modelling software. Third, we conduct three case studies applying the proposed framework to real-world scientific models from different domains, evaluating its ability to predict metamorphic test outcomes from observational data.

The remainder of this paper is structured as follows. \Cref{sec:background} provides a motivating example and necessary background. \Cref{sec:causal-testing-framework} introduces our conceptual framework that frames causality-driven testing activities as problems of CI. \Cref{sec:reference-implementation} then introduces our reference implementation of this framework, before demonstrating its application to three real-world scientific models in \Cref{sec:case-studies} and discussing the main findings and threats to validity in \Cref{sec:discussion}. \Cref{sec:related-work} reviews related work, and \Cref{sec:conclusion} concludes the paper.

\section{Background and Preliminaries}\label{sec:background}
This section defines the scope of the paper and introduces the main challenges associated with testing scientific modelling software, as outlined in Kanewala and Bieman's survey on the same topic \cite{kanewala2014testing}. We present these challenges in the context of a real-world, motivating example that is used as one of three case studies in \Cref{sec:case-studies}. We then provide a background on model-based testing and, in particular, metamorphic testing \cite{chen1998metamorphic}, a known solution to some of these challenges. We conclude this section with a brief introduction to causal inference, the statistical methodology employed by the framework presented in \Cref{sec:causal-testing-framework}.

\subsection{Black-Box Software Systems}
In this paper, we view and test software from a black-box perspective \cite{nidhra2012black}, focusing on the relationships between its inputs and outputs rather than its inner-workings and source code. More formally, in this paper, we conceptualise the system-under-test (SUT) as follows:

\begin{definition}\label{def:system-under-test}
A \emph{system-under-test} (SUT) is a software system comprising a set of input variables, $I$, and output variables, $O$, such that $I \cap O = \oslash$. We consider inputs to be parameters whose values are set prior to execution that influence the resulting system behaviour. We consider outputs to be features of the system that can be measured at any point during or after execution without inspecting or modifying the source code.
\end{definition}

Given our focus on causality in this paper, we provide an informal definition of causality in \Cref{def:causality}. This follows from Pearl's characterisation of causation, which states that ``variables earn causal character through their capacity to sense and respond to changes in other variables'' \cite{pearl2018does}.

\begin{definition}\label{def:causality}
We say that a variable $X=x$ \emph{causes} a variable $Y$ if there exists some value $x'$ such that, had the value of $X$ been changed to $x'$, the value of $Y$ would change in response. 
\end{definition}

Furthermore, we are primarily interested in scientific modelling software. Informally, we consider this to be any form of software that has a significant computational component and simulates, models, or predicts the behaviour of complex, uncertain phenomena to support policy and scientific decisions \cite{kanewala2014testing, kreyman1999inspection}. We focus on this form of software as it typically possesses a number of challenging characteristics that preclude the application of many conventional testing techniques, but can be addressed by the framework introduced in \Cref{sec:causal-testing-framework}. In the following section, we introduce a motivating example to familiarise the reader with these challenging properties.

\subsection{Motivating Example: Covasim}\label{sec:motivating-example}
Covasim \cite{kerr2021covasim, covasim2022github} is an epidemiological agent-based model that has been used to inform COVID-19 policy decisions in several countries \cite{kerr2021controlling, cohen2020schools, panovska2020determining, scott2020modelling}. Given the critical applications of such scientific models, it is of paramount importance that they are tested to the best of our abilities. However, Covasim has a number of characteristics that make testing particularly challenging.

Covasim has a \textbf{vast and complex input space}, with 64 unique input parameters, 27 of which are complex objects characterised by further parameters. Furthermore, the \textbf{precise values for many of the inputs are unknown} and are instead described by a distribution, meaning that any given scenario can be simulated using a potentially intractable number of input configurations.

Covasim also suffers from \textbf{long execution times and high computational costs}. Non-trivial runs of Covasim can take hours and accumulate large amounts of data. To compound this issue further, the model is also \textbf{non-deterministic}: running the same simulation parameters multiple times (with a different seed) will yield different results, meaning that each modelling scenario must be simulated several times to observe a distribution of outcomes.

Additionally, Covasim encounters the oracle problem: for most modelling scenarios, \textbf{the precise expected output is unknown}. This makes Covasim a traditionally ``untestable'' \cite{weyuker1982} system as it is difficult to determine whether the output of a given test is correct.

Despite these challenges, Covasim features a mixture of unit, integration, and regression tests achieving 88\% code coverage\footnote{Code coverage obtained from commit \href{https://github.com/InstituteforDiseaseModeling/covasim/commit/7da3bc46e2344fa8128dfa66c260cadf4213bea2}{7da3bc4}.}. However, many of these tests lack a test oracle and appear to rely on the user to determine correctness instead. For example, the vaccine intervention has two tests \cite{covasim2022githubVaccineTests} that instantiate and run the model with two different vaccines and plot the resulting model outputs on a graph for manual inspection.

While the existing vaccination tests reveal the difference in outcome \emph{caused} by changing from one vaccine to another, the experimental approach employed would not scale well if the tester wanted to test more general properties that cover larger value ranges. For example, tests covering multiple versions of vaccine (Pfizer, Moderna, etc.) and outcomes (infections, hospitalisations, etc.). However, this is not a criticism of Covasim, but a statement that conventional testing techniques are impractical for testing scientific modelling software. Hence, there is a clear need for testing techniques more sympathetic to their challenging characteristics.

\subsection{Model-Based Testing}


An approach that is often used to test black-box systems is model-based testing \cite{2004test}. The main principle behind model-based testing is to provide a model that captures the expected behaviour of the SUT \cite{utting2010practical}.
Such a model incorporates invaluable domain expertise and can form the basis for test generation, with work in this area going back to the 1950s \cite{moore56}.
In addition, if the model has formal semantics, testing can be represented as a process in which one compares the behaviour of two models: the known specification model $M$ and an unknown model $N$ that represents the behaviour of the SUT.
It is then possible to reason about the effectiveness of testing \cite{gaudel95,Tretmans08}.
Note that since a model describes the expected behaviour of the SUT, it can also form the basis of a test oracle,
and this is at least implicit in most MBT work \cite{gaudel95,Tretmans08}.

For testing black-box systems (i.e. where the internal workings are unknown to the test developer), an appropriate model will typically specify formal relations between the inputs and outputs of the SUT. For example, pre/post models can be defined in various modelling languages, such as Z \cite{spivey92} and B \cite{ButlerKKLLMV20}, that model a system as a collection of variables and captures the expected behaviour in terms of pairs of pre-conditions and post-conditions \cite{utting2010practical}. In this way, testers use their domain expertise to specify how they expect the SUT to respond under different settings.

However, for complex software like Covasim that suffers from the test oracle problem \cite{barr2015}, it is seldom possible to specify the expected outputs or post-conditions corresponding to a particular set of inputs or pre-conditions. As discussed in \Cref{sec:motivating-example}, this is partly due to the exploratory nature of Covasim that makes it difficult (if not impossible) to establish what `correctness' looks like. This is typically the case for any form of scientific software primarily used to predict or simulate future events, such as meteorological software for predicting the weather.
Under such circumstances, the domain expertise needed to specify a model of the expected behaviour are fundamentally unattainable, preventing the tester from capturing static input-output relations, such as pre/post models, a priori. 

One solution that effectively avoids the oracle problem and has been advocated as a technique for testing scientific software \cite{kanewala2014testing} is \emph{metamorphic testing} \cite{chen1998metamorphic}. The basic idea is to model the expected behaviour of the SUT as so-called \emph{metamorphic relations} that describe the expected change in output in response to a specific \emph{change} in input. For example, to test an implementation of $\sin$, we may assert that $\forall x.\,\sin(x) = \sin(\pi-x)$. These relations provide a means of generating test cases and validating the observed behaviour \cite{segura2016survey}. By stating the expected behaviour in terms of \emph{changes} to inputs and outputs, we can test the system without knowing the precise expected outcome corresponding to some inputs.

Statistical metamorphic testing (SMT) \cite{guderlei2007statistical} generalises this to non-deterministic systems, which produce different outputs when run repeatedly under identical input configurations. Rather than comparing outputs directly, the SUT is run multiple times for each input configuration and statistical tests are performed on the corresponding distributions of outputs. However, the potentially high computational costs involved in this process are a major limitation to the applicability of SMT to scientific models.


\subsection{Causal Inference}
\label{sec:causal-inference}
The framework we present in \Cref{sec:causal-testing-framework} uses a family of statistical techniques, known as causal inference (CI), designed to make claims about causal relationships between variables \cite{keele2015statistics}. Our goal is to use this family of techniques to provide an efficient method for testing scientific software. Here we provide a brief introduction to the essential notions of CI used in this work. For a more comprehensive overview, we refer the reader to \cite{pearl2009causal, hernan2020causal}.

\subsubsection{Preliminaries}
Causality is often presented in terms of the ``ladder of causality'' \cite{pearl2018why}, which groups different tasks into three `rungs': Rung one is \emph{observation and association} as per traditional statistical methods; Rung two is \emph{intervention}, which imagines the effects of taking particular actions: ``What if I do...?'', and rung three is \emph{counterfactual}, which imagines the effects of retrospective actions: ``What if I had done...?''.

Traditional statistical approaches are limited to rung one. By simply observing the association between variables (in our case input and output variables), without systematically controlling the selection of values or resorting to additional domain knowledge, it is impossible to answer fundamentally \emph{causal} questions \cite{pearl2009causal}. This problem is commonly captured by the adage: ``correlation does not imply causation''.

CI enables us to estimate and quantify causal effects in order to make claims about causal relationships \cite{keele2015statistics}. Informally, the causal effect of a treatment $T$ on an outcome $Y$ is the change in $Y$ that is caused by a specific change in $T$ \cite{pearl2018why}. In this context, a \emph{treatment} is a variable that represents a particular action or intervention, such as changing a line of code, and an \emph{outcome} is an observable feature or event, such as the occurrence of a fault.

One of the main challenges underlying CI is the design of experiments or statistical procedures that mitigate sources of bias to isolate and measure causality (rungs two and three) as opposed to association (rung one). In fields such as medicine, randomised control trials (RCTs) are often considered as the gold standard approach for CI \cite{cartwright2010randomised}. RCTs mitigate sources of bias by randomly assigning subjects to either the treatment or control group \cite{kendall2003designing}. However, there are many situations in which RCTs cannot be performed due to ethical or practical reasons \cite{adebamowo2014randomised}.

Where RCTs cannot be performed, researchers often turn to observational data and statistical models as means for conducting CI. At a high level, this observational approach to CI can be broken down into two tasks: identification and estimation. Identification involves identifying sources of bias that must be adjusted for statistically in order to obtain a causal estimate. Estimation is the process of using statistical estimators, adjusted for the identified biasing variables, to estimate the causal effect.

\subsubsection{Metrics}
Several metrics can be used to measure causal effects. Perhaps the most desirable is the \emph{individual treatment effect} (ITE), which describes the effect of a given treatment on a particular individual. In the majority of cases, however, individual-level inferences are unattainable due to the \emph{fundamental problem of causal inference} \cite{holland1986statistics}; namely that, for a given individual, it is usually only possible to observe the outcome of a single version of treatment (e.g. an individual either takes an aspirin for their headache or does not).

To address this, researchers typically turn to population-level causal metrics, such as the \emph{Average Treatment Effect} (ATE) \cite{hernan2020causal}:
$$\text{ATE} = \sum_{z \in Z}\expe{[Y \mid X=x_t, Z=z]} P(Z=z) - \sum_{z \in Z} \expe{[Y \mid X=x_c, Z=z]} P(Z=z)$$
The ATE quantifies the average additive change in outcome we expect to observe in response to changing some treatment variable $X$ from the \emph{control value} $x_c$ to the \emph{treatment value} $x_t$, while adjusting for all biasing variables $Z$. However, in some instances, it is desirable to refine our inferences to specific sub-populations defined by some notable characteristic. To this end, the conditional ATE (CATE) can be obtained by applying the ATE to specific sub-populations of interest \cite{abrevaya2015estimating}.



An alternative causal metric is the \emph{Risk Ratio} (RR) \cite{hernan2020causal}:

$$\text{RR} = \frac{\sum_{z \in Z} \expe{[Y \mid X=x_t, Z=z]} P(Z=z)}{\sum_{z \in Z} \expe{[Y \mid X=x_c, Z=z]}P(Z=z)}$$

The RR captures the multiplicative change in an outcome $Y$ caused by changing the treatment variable $X$ from the control value $x_c$ to the treatment value $x_t$ while adjusting for all biasing variables $Z$.

Other effect metrics such as the \emph{odds ratio} (OR) and the \emph{effect of treatment on the treated} (ATT) also exist but fall outside the scope of this paper. Furthermore, to quantify uncertainty, effect measures are typically accompanied by 95\% confidence intervals that quantify the interval within which we are 95\% confident the true estimate lies \cite{o2016interpret}.

\subsection{Causal DAGs}
\label{sec:casualDAGs}
CI generally depends on domain expertise and causal assumptions that cannot be tested in practice \cite{rothman2005causation}. Given that different domain experts may make different assumptions about the same problem and that these may lead to different results, it is essential that all assumptions are made transparent. To this end, causal DAGs provide an intuitive graphical method for communicating the causal assumptions necessary to solve CI problems \cite{pearl1995}. Formally, a causal DAG is defined as follows \cite{hernan2020causal}:

\begin{definition}
    A causal DAG $G$ is a directed acyclic graph (DAG) $G=(V, E)$ comprising a set of nodes representing random variables, $V$, and a series of edges, $E$, representing causality between these variables, where:
    \begin{enumerate}
        \item The presence/absence of an edge $V_i \to V_j$ represents the presence/absence of a direct causal effect of $V_i$ on $V_j$.
        \item All common causes of any pair of variables on the graph are themselves present on the graph.
    \end{enumerate}
    \label{def:causalGraph}
\end{definition}

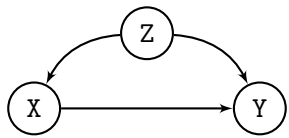
\begin{figure}[ht!]
    \centering
    \begin{tikzpicture}[every node/.style={circle,draw}]
        \node[vertex] (X) at (0, 0) {$\mathtt{X}$};
        \node[vertex] (Y) at (3, 0) {$\mathtt{Y}$};
        \node[vertex] (Z) at (1.5, 1) {$\mathtt{Z}$};
        \draw[edge] (X) to (Y);
        \draw[edge] (Z) to [bend right] (X);
        \draw[edge] (Z) to [bend left] (Y);
    \end{tikzpicture}
    \caption{An example causal DAG for the causal effect of $\mathtt{X}$ on $\mathtt{Y}$ confounded by $\mathtt{Z}$.}
    \label{fig:causalDiagram}
\end{figure}

In \Cref{fig:causalDiagram}, \circled{X}, \circled{Y}, and \circled{Z} are nodes representing \emph{random variables}, which, in this context, are variables that can take different values for different individuals (e.g. people or software executions). We say that $\mathtt{X}$ is a \emph{direct cause} of $\mathtt{Y}$ because there is an edge from $\mathtt{X}$ directly into $\mathtt{Y}$. We refer to $\mathtt{Y}$ as a \emph{descendant} of $\mathtt{Z}$ and $\mathtt{X}$ because there is a sequence of edges, known as a \emph{path}, such that, if you follow the direction of those edges, you can reach $\mathtt{Y}$ from $\mathtt{Z}$. That is, $\mathtt{Z} \to \mathtt{X} \to \mathtt{Y}$.

As mentioned in the previous section, in order to estimate the causal effect of $\mathtt{X}$ on $\mathtt{Y}$, we need to identify and adjust for all variables that bias the relationship $\mathtt{X \to Y}$. Using a causal DAG, we can achieve this automatically by applying a pair of graphical tests, the \emph{back-door criterion} and \emph{d-separation}, which are formally defined as follows:

\begin{definition}
A path $p$ is \emph{blocked} or \emph{d-separated} by a set of variables $Z$ if and only if at least one of the following conditions hold \cite{pearl2009causality}:
\begin{enumerate}
    \item $p$ contains a chain $i \to k \to j$ or a fork $i \leftarrow k \to j$ where $k \in Z$.
    \item $p$ contains a collider $i \to k \leftarrow j$ where $k \notin Z$ and for all descendants $k'$ of $k$, $k' \notin Z$.
\end{enumerate}
\label{def:block}
\end{definition}

\begin{definition}
A set of variables $Z$ is said to satisfy the \emph{back-door criterion} relative to an ordered pair of variables $(X, Y)$ if both of the following conditions hold \cite{pearl2009causality}:
    \begin{enumerate}
    \item No variable in $Z$ is a descendant of $X$.
    \item Z blocks every path between $X$ and $Y$ that contains an arrow into $X$.
\end{enumerate}
\label{def:backDoorCriterion}
\end{definition}

A set of variables $Z$ is said to be a \emph{sufficient adjustment set} relative to a pair of variables ($X$, $Y$) if adjusting for $Z$ blocks all back-door paths between $X$ and $Y$. Conceptually, this corresponds to a set of variables that, once adjusted for, mitigate all known sources of bias and that is therefore capable of isolating the \emph{causal effect} of interest. For example, in \Cref{fig:causalDiagram}, $\mathtt{Z}$ satisfies the back-door criterion relative to $\mathtt{\left(X, Y\right)}$ because $\mathtt{Z}$ blocks every path between $\mathtt{X}$ and $\mathtt{Y}$ with an arrow into $\mathtt{X}$. Therefore, we can endow the ATE of $\mathtt{X}$ on $\mathtt{Y}$ with a causal interpretation and estimate its value directly using the following closed-form statistical expression:
$$\sum_{\mathtt{z \in Z}} \expe{[\mathtt{Y \mid X=1,Z=z}]} P(\mathtt{Z=z}) - \sum_{\mathtt{z \in Z}} \expe{[\mathtt{Y \mid X=0, Z=z}]} P(\mathtt{Z=z})$$



Overall, causal DAGs provide a principled and automated approach for designing statistical `recipes' capable of measuring causal relationships and endowing statistical measures with causal interpretations. In the following section, we introduce a framework that facilitates the application of this approach to the testing of scientific modelling software. Furthermore, we opt to use graphical CI over other CI frameworks, such as potential outcomes \cite{rubin2005causal} or structural equation modelling \cite{kline2015principles}, as it provides a transparent and intuitive way to both specify and test causal relationships, without necessarily requiring users to know their precise functional form.

\section{Causal Testing Framework}\label{sec:causal-testing-framework}
This section introduces the Causal Testing 
Framework (CTF): a conceptual framework that approaches causality-driven testing activities as CI problems. That is, testing activities that intend to establish the (inherently causal) relationship between inputs and outputs, such as metamorphic testing. By framing testing activities in this way, it is possible to leverage CI techniques to make strong claims about causal relationships between inputs and outputs, and to do so in an efficient manner by exploiting data from previous test executions.

In the remainder of this section, we define four key components of our causal testing framework: specifications, programs, tests, and oracles~\cite{staats2011programs}, giving an example using Covasim (see \Cref{sec:background}) for each component. We also provide informal guidance for constructing causal DAGs and examine the relationship between the CTF and metamorphic testing.

\subsection{Causal Specification}
 In the CTF, our primary aim is to test scientific models in terms of the effects of interventions. Given the diverse range of possible scenarios that a typical scientific model can simulate, we further focus on testing individual modelling scenarios. We define a modelling scenario as a series of constraints placed over a subset of the SUT's (see \Cref{def:system-under-test}) input variables that characterise the scenario of interest. Therefore, in the causal testing framework, the set of programs are programs that implement modelling scenarios $\mathcal{M}$ (\Cref{def:modelling-scenario}).

\begin{definition}\label{def:modelling-scenario}
A \emph{modelling scenario} $\mathcal{M}$ is a pair $(X, C)$ where $X$ is a non-strict subset of the model's input variables and $C$ is a set of constraints over realisations of $X$, which may be empty.
\end{definition}

The expected behaviour of scientific modelling software in a given scenario depends on a series of underlying modelling assumptions. It is therefore essential that such modelling assumptions are made transparent and readily available, particularly for the purposes of testing. Indeed, past investigations into modelling failures have highlighted the importance of transparency and accountability \cite{oldfield2021analytical}. In the same vein, causal testing requires an explicit record of causal assumptions to enable the transparent and reproducible application of graphical CI techniques. To this end, we use a causal DAG that captures causality amongst a subset of the SUT's input and outputs. Therefore, we define a \textit{causal specification} (\Cref{def:causal-specification}) as a pair comprising a modelling scenario ($\mathcal{M}$) and a causal DAG ($\mathcal{G}$).

\begin{definition}\label{def:causal-specification}
  A \emph{causal specification} is a pair $\mathcal{S} = (\mathcal{M}, \mathcal{G})$ comprising a modelling scenario $\mathcal{M}$ and a causal DAG $\mathcal{G}$ capturing the causal relationships amongst the inputs and outputs of the SUT that are central to the modelling scenario.
\end{definition}

\begin{example}\label{ex:causal-spec}
  Consider a scenario in Covasim where we want to test the effect of prioritising the elderly for vaccination $V$ on the total vaccine doses administered $N_D$, total vaccinated agents $N_V$, maximum number of doses per agent $M_D$, and cumulative infections $I$.
  Further, let us restrict our simulation length to 50 days, the initial number of infected agents to 1000, and the population size to 50,000.
Our modelling scenario is then characterised by the constraints $\{ \mathtt{days}=50, \mathtt{pop\_size}=50000, \mathtt{pop\_infected}=1000\}$, and the causal DAG is the set of edges $\{V \to N_V, V \to N_D, V \to I\}$. Note the absence of edge $V \to M_D$. Here we are asserting that $V$ may cause a change in $N_V$, $N_D$, and $I$, but should cause no change to $M_D$. This is because at most two doses of the vaccine are administered to each at agent so changing the target population should not affect this.
\end{example}

\subsection{Constructing Causal DAGs}
\label{sec:dag-guidelines}
In the testing context, causal DAGs offer a flexible, lightweight means by which to capture potential causal relationships between inputs and outputs. Here we present a set of guidelines for constructing the graph (informed by our experience with the case studies).

We start by constructing a complete directed graph over the set of inputs and output: $I \cup O$. Then, to simplify this structure, we apply the following assumption:

\begin{assumption}
    \label{assumption:assumption1}
    Outputs cannot cause inputs.
\end{assumption}

\Cref{assumption:assumption1} follows from temporal precedence (that a cause must precede its effect) \cite{pearl1995theory} and the observation that, in a given test execution, outputs temporally succeed inputs. This enables us to delete all edges from outputs to inputs.

Then, in many cases, we can also apply the following assumption to remove all edges from inputs to inputs:

\begin{assumption}
    \label{assumption:assumption2}
    Inputs cannot cause changes to the values of other inputs and, therefore, cannot share causal relationships.
\end{assumption}

As stated in \Cref{def:system-under-test}, in this paper, we assume that all inputs are assigned their values prior to execution. Under this characterisation, changes to the value of one input cannot \emph{physically} affect another input's value and, therefore, inputs cannot share causal relationships. Of course, there are caveats to this; if a system has input validation, for example, the assignment of one input's value may \emph{physically} restrict which values can be selected for a second input. Note that, in such cases, our framework is still applicable, but the user would have to consider more edges manually to construct their DAG.


This leaves us with the following forms of potential causal relationships to consider: $I \to O$ and $O \to O$ (and $I \to I$ if \Cref{assumption:assumption2} cannot be applied). Output to output causality may occur in software where an earlier output is used in the computation of a later output. For example, in a weather forecasting model, a prediction of the weather in three days time is affected by the weather predicted for one and two days time.

This is the point at which the tester's domain knowledge is fed into the model, by pruning edges where they are certain that there is no causal relationship (see \Cref{def:causality} for an informal definition of causality). We recommend following this approach of pruning edges from a complete directed graph over adding edges to an initially empty graph, as the absence of an edge carries a stronger assumption than the presence of one \cite{tennant2021use}. This follows from the fact that the presence of an edge states that there exists \emph{some} causal relationship, whereas the absence of an edge states that there is \emph{precisely} no causal relationship.

\subsection{Causal Testing}
Causal testing draws its main inspiration from CI, which focuses on the effects of \emph{interventions} on \emph{outcomes}. In this context, an intervention manipulates an input configuration in a way that is expected to \emph{cause} a specific outcome to change. Here, we refer to the pre-intervention input configuration as a \emph{control} and the post-intervention input configuration as a \emph{treatment}. A causal test case then specifies the expected change in outcome caused by this intervention (i.e. the expected causal effect). When phrased this way, causal tests bear a remarkable similarity to metamorphic tests, highlighting the fact that, at its core, metamorphic testing can be viewed as an inherently a causal activity. We explain this relationship further in Section~\ref{sec:metamorphic-relationship}.

\begin{definition}\label{def:intervention}
  An \emph{intervention} $\Delta :\mathcal{X} \to \mathcal{X'}$ is a function which manipulates the values of a subset of input realisations.
\end{definition}

\begin{definition}\label{def:causal-test-case}
  A \emph{causal test case} $\mathcal{T}$ is a 4-tuple $(\mathcal{M}, \mathcal{X}, \Delta, \mathcal{Y})$ that captures the expected causal effect, $\mathcal{Y}$, of an intervention, $\Delta$, made to an input valuation, $\mathcal{X}$, on some model outcome in the context of modelling scenario $\mathcal{M}$. The expected causal effect $\mathcal{Y}$ is an informal expression of some change in outcome that is expected to be caused by executing $\mathcal{T}$. We refer to the input realisation $\mathcal{X}$ as the control input configuration.
\end{definition}

\begin{example}
  Continuing with our vaccination example, suppose we want to create a causal test case that investigates the effect of switching vaccine from $\mathtt{Pfizer}$ to an age-restricted version ($\mathtt{Pfizer'}$) on only the maximum number of doses per agent $M_D$. We can start by using the modelling scenario outlined in the previous example and then specify our control input configuration as the input realisation $\mathcal{X} = \mathtt{\{vaccine=Pfizer\}}$. We then define an intervention that takes the control input configuration and replaces the vaccine with the age-restricted version: $\Delta(\mathcal{X}) = \mathcal{X}[\mathtt{vaccine} := \mathtt{Pfizer'}]$. We complete our causal test case by specifying the expected causal effect, $\mathcal{Y}$: the intervention should cause no change to $M_D$ and we therefore expect that the ATE will be zero.
\end{example}

Finally, we must consider the test oracle: the \emph{procedure} used to determine whether the outcome of a causal test case ($\mathcal{T}$) is correct (i.e. whether it realises the expected causal effect $\mathcal{Y}$). In the context of causal testing, the oracle must ascertain the correctness of causal estimates relative to a modelling scenario ($\mathcal{M}$). Therefore, we refer to our oracle as a causal test oracle (\Cref{def:intervention}).

\begin{definition}\label{def:causal-test-oracle} A \emph{causal test oracle} $\mathcal{O}$ is a procedure, such as an assertion, that determines whether the outcome of a causal test case $\mathcal{T}$ is correct or incorrect. This procedure checks whether the application of the intervention $\Delta$ to the control input configuration $\mathcal{X}$ has caused the expected causal effect $\mathcal{Y}$ in the context of modelling scenario $\mathcal{M}$.





\end{definition}

\begin{example}
  Continuing with our Covasim example, for the causal test case $\mathcal{T}$ defined in the previous example, our causal test oracle must check whether applying the intervention (i.e. replacing the $\mathtt{Pfizer}$ vaccine with an age-restricted version $\mathtt{Pfizer'}$) has no effect on $M_D$, as specified by the expected causal effect $\mathcal{Y}$. We can implement this test oracle as the following assertion: $\mathtt{ATE_{M_D} = 0}$. This checks whether the change in $M_D$ caused by the intervention ($ATE_{M_D}$) is zero, as expected.
  
\end{example}

Notice the subtle difference between the expected causal effect, $\mathcal{Y}$, of the causal test case, $\mathcal{T}$, and the causal test oracle, $\mathcal{O}$: the former is a statement of the \emph{expected test outcome} while the latter is the \emph{actual procedure} used to check whether the anticipated outcome holds. We make this distinction with the transparency of the causal testing process in mind, avoiding situations where two testers may implement the procedure to ascertain the validity of a given causal test case in different ways, potentially leading to different test outcomes. In other words, the CTF considers the expected outcome ($\mathcal{Y}$) and the procedure used to check this has been realised ($\mathcal{O}$) as separate entities that carry equal importance.


Any discrepancy between the test result and the expected outcome revealed by the test oracle implies one of two problems: (i) the implementation contains a bug or an error, or (ii) the underlying causal knowledge is incorrect. It follows that causal testing lends itself to an iterative testing process \cite{murphy1995software}, whereby the user inspects the source code to explain any identified discrepancies and, if no bugs are found, reviews the causal DAG to check if the underlying science is correct.

Collectively, the components of the CTF enable the application of graphical CI techniques to testing activities that concern the causal effect of some intervention. In theory, the CTF should therefore provide the following advantages over existing solutions:
\begin{compactenum}
    \item The ability to derive test outcomes \emph{experimentally}\footnote{We use the term `experimental' loosely here; the CTF performs a quasi-experiment in which the SUT is executed with a pair of input configurations that isolate the causal effect of the intervention on the output. Specifically, the SUT is executed twice: once using the pre-intervention configuration and once using the post-intervention configuration. This is repeated multiple times for non-deterministic systems.} (by strategic model executions that isolate a particular cause-effect relationship by design) and \emph{observationally} (by applying CI techniques to past execution data).
    \item The ability to identify and adjust for confounding bias in observational data using a causal DAG.
    From a testing perspective, this effectively relaxes the experimental conditions ordinarily required to reach causal conclusions. Namely, the need for carefully controlled, unbiased test data.
    \item The ability to derive \emph{counterfactual} test outcomes using appropriate statistical models.
    This would enable testers to infer how the model would likely behave, had it been run under a different parameterisation. Therefore, where practical constraints preclude further executions of the SUT, counterfactual inference can offer a cost-effective alternative.
\end{compactenum}



In \Cref{sec:case-studies}, we apply the CTF to a series of real-world scientific models to understand how a modeller can leverage these advantages in a testing context to improve the efficiency and applicability of metamorphic testing; a state-of-the-art approach for testing scientific modelling software. 


\subsection{Relationship to Metamorphic Testing}\label{sec:metamorphic-relationship}
At a high level, the CTF and metamorphic testing share the same objective: to evaluate the \emph{effect} caused by making a change to some input.

Metamorphic testing provides a means of generating ``follow-up test cases'' using metamorphic relations which should hold over a number of different parameter values \cite{barr2015,segura2016survey}. In contrast to typical program invariants, which must hold for every execution of a given program, metamorphic relations hold between different executions. In other words, they investigate the effect of a change (or \emph{intervention} in causal language) on an input. This is a key similarity between causal testing and metamorphic testing.

In this sense, metamorphic tests can be thought of as quasi-experiments\footnote{We liken metamorphic tests to quasi-experiments rather than controlled experiments as they lack an explicit randomisation mechanism.} designed to answer causal questions about the SUT. For example, a metamorphic test for our property of the $\sin$ function in \Cref{sec:background} that $\forall x.\,\sin(x) = \sin(\pi-x)$ can be thought of as a quasi-experiment that confirms whether changing the input from $X=x$ to $X=\pi-x$ causes no change to the output. That is, there should be \emph{no causal effect}. This synergism suggests that metamorphic testing can be re-framed and solved as a problem of CI and, therefore, benefit from its advantages. To this end, in \Cref{sec:case-studies}, we demonstrate how the CTF can conduct metamorphic testing using CI techniques.

One advantage of causal testing over traditional metamorphic testing is that causal testing does not necessarily require dedicated test runs of the system to be performed if sufficient test data already exists. Even (and especially) if this data is biased, CI can account for this, meaning that testing can be performed on systems which cannot be tested for reasons of practicality. Furthermore, systems can be tested retroactively, enabling concerns about a model's correctness to be investigated even after the model has been run. This is potentially advantageous in the context of scientific models, where their integrity and correctness can be called into question years after policies based on their output have already been made. In such situations, the DAG makes clear the assumptions made about the functionality of the model so adds weight to any conclusions made.

\section{CTF Reference Implementation}\label{sec:reference-implementation}
This section provides an overview of our open-source Python reference implementation of the Causal Testing Framework (CTF)\footnote{\url{https://github.com/CITCOM-project/CausalTestingFramework}}, comprising over 4000 lines of Python code, and outlines four stages of the CTF workflow: Specification, Test Cases, Data Collection, and Testing.

\subsection{Causal Specification}\label{sec:causal-specification}
To begin causal testing, we form a causal specification (\Cref{def:causal-specification}), comprising two components: a modelling scenario and a causal DAG. We form the modelling scenario by specifying a set of constraints over the inputs that characterise the scenario-under-test, such as $x_1 < x_2$. Next, we specify our causal DAG using the DOT language \cite{ellson2002}, in which graphs are expressed as a series of edges, such as $x_1 \to x_2$, following the guidelines outlined in \Cref{sec:dag-guidelines}.

\subsection{Causal Test Case}\label{sec:causal-test-case}
Now that we have a causal specification, we define a causal test case that describes the intervention whose effect we wish to test. In our reference implementation, a causal test case is an object that requires us to specify a control input configuration, a treatment input configuration, and the expected causal effect. In the following steps, this information will enable us to collect appropriate test data (Data Collection), design quasi-experiments isolating the causal effect of interest within this data, and define test oracles that ascertain whether the expected causal effect is observed (Causal Testing).

\subsection{Data Collection}\label{sec:data-collection}
After creating a causal specification and causal test case, the next step is to collect data corresponding to the modelling scenario. We can achieve this either (quasi-)experimentally (in situations where we are able to directly execute the SUT) or observationally (in situations were we are not able to execute the SUT, but are instead able to draw upon prior execution data).

\subsubsection{Experimental Data Collection}
Experimental data collection executes the model \emph{directly} under both the control and treatment input configuration to isolate the causal effect of the intervention. To this end, our reference implementation provides an abstract experimental data collector class, requiring us to implement one method that executes our model with a given input configuration. This method enables the CTF to run the model under the conditions necessary to isolate causality directly.

\subsubsection{Observational Data Collection}
Since it is often infeasible to run models a statistically significant number of times, we also provide the option to use observational, existing test data. This data may not meet the experimental conditions necessary to isolate the causal effect and thus may contain biases that lead purely statistical techniques astray. However, by employing graphical CI techniques, the CTF can identify and mitigate bias in the data, providing an efficient method for testing scientific models \emph{a posteriori}.

There are two caveats to this. First, the causal DAG must be correctly specified. While this is not generally verifiable, several techniques exist that can quantify the sensitivity of casual estimates to unobserved confounding, including the robustness value \cite{cinelli2020making} and the e-value \cite{vanderweele2017sensitivity}. These techniques could be employed to justify that the DAG-informed adjustment set yields causal estimates that are robust to missing confounders. Second, the observational data must be consistent with the constraints of the causal specification. To this end, our reference implementation includes an observational data collector class that takes a CSV file of existing test data as input and uses the Z3 theorem prover \cite{deMoura2008} to identify and remove any executions of the SUT that violate constraints. By execution, we refer to an individual run of the SUT with some set of inputs that produces some set of outputs. We assume the CSV file comprises a row for each such execution, with a column for each input and output value. Next, we describe how the CTF infers test outcomes from this data.

\subsection{Causal Testing}
Given a causal test case, testing is carried out in two stages: causal inference (CI) and applying the test oracle.

\subsubsection{Causal Inference} To infer the causal effect of interest, our reference implementation applies the two steps of CI outlined in \Cref{sec:background}: identification and estimation. For identification, the CTF algorithmically identifies an adjustment set (see \Cref{sec:causal-inference}) for the causal effect of interest. Then, for estimation, we design an appropriate estimator that adjusts for the identified adjustment set, and apply the estimator to our data to estimate the desired causal metric (e.g. ATE or RR, see \Cref{sec:background}). To this end, our reference implementation provides regression and causal forest \cite{wager2018estimation} estimators which can be customised to add additional features such as squared and inverse terms to change the shape of the model. In addition, the CTF includes an abstract estimator class that enables users to define their own estimators. This step outputs a causal test result containing the inferred causal estimate for the desired causal metric (e.g. ATE or RR, see \Cref{sec:causal-inference}) and 95\% confidence intervals. The user is, of course, free to relax their confidence intervals should they wish to obtain a more precise estimate with a higher level of associated risk, or vice versa.

\subsubsection{Test Oracle}
After applying CI, all that remains is the test oracle procedure. That is, to check whether the causal test results match our expectations. For this purpose, our reference implementation provides several test oracles that check for positive, negative, zero, and exact effects. Alternatively, to handle more complex outputs, a user can specify a custom oracle that ascertains whether a causal test result should pass or fail.

Now that we have discussed the workflow of our CTF reference implementation, in the following section, we demonstrate its application to three vastly different real-world scientific models.

\section{Case Studies}\label{sec:case-studies}
This section applies the Causal Testing Framework (CTF) to four testing scenarios covering three real-world scientific models from different domains, approaching (statistical) metamorphic testing as a CI problem. Our goal here is to conduct a series of \emph{evaluative} case studies \cite{ralph2020acm} that appraise the CTF with respect to three attributes: \emph{accuracy}, \emph{efficiency}, and \emph{practicality}. Here, we do not aim to make generalisable conclusions, but to evaluate the CTF with respect to each of these attributes within the context of each subject system. To this end, across our case studies, we corroborate evidence to collectively answer the following research questions:

\paragraph{\textbf{RQ1 (Accuracy): Can we reproduce the results of a conventional MT/SMT approach by applying the CTF to observational data?}}
As mentioned in \Cref{sec:introduction}, CI is a generally applicable technique \cite{bareinboim2016} promising the ability to infer test outcomes from existing data that is potentially confounded. In the context of testing scientific software, this approach has the potential to reduce the overhead associated with SMT by enabling the inference of metamorphic test outcomes from existing execution data. This is in contrast to a conventional approach which may require numerous potentially costly executions.

In this research question, we consider whether the CTF is able to predict metamorphic test outcomes from observational data with sufficient accuracy to make \emph{actionable inferences}. By actionable inferences, we refer to predicted outcomes that provide a truthful and meaningful insight into the actual behaviour of the SUT.

\paragraph{\textbf{RQ2 (Efficiency): In terms of the amount of data required, is the CTF more cost effective than a conventional MT/SMT approach?}}
In practice, the utility and applicability of the CTF depends on the amount of observational data required to make actionable inferences. Hence, for the CTF to be considered a useful tool and a viable alternative to conventional MT and SMT approaches, it must be capable of making actionable inferences using no more data than is required by a conventional approach. 

To this end, in order to understand the efficiency and therefore utility of the proposed approach, this research question investigates the relationship between the amount of observational data and the accuracy of insights provided by the inferred metamorphic test outcomes.


\paragraph{\textbf{RQ3 (Practicality): What practical effort is required from the tester to conduct MT/SMT using the CTF?}}
The CTF requires causal knowledge and domain expertise that, in turn, depend on human effort. This human effort cannot be overlooked. Hence, in order to determine whether the technique can be considered practical and applicable, it is necessary to investigate the trade-off between the human cost and the benefits offered by the CTF.

In this research question, we provide a qualitative account of the human effort involved in applying the CTF to each case study. \\


In the remainder of this section, we cover each of the three case studies in accordance to the following high-level structure. 
First, we describe the characteristics of the subject system and our justification for selecting it. We then provide a brief overview of the testing activity (the broad testing objective) and the process of acquiring data for analysis. Following this, we describe the application of the CTF and analyse the generated data. We conclude by analysing the outcomes and answering the relevant research questions. The contribution of each case study to the research questions will be highlighted throughout the case studies and the collective findings will be discussed in \Cref{sec:discussion}.

\subsection{Poisson Line Tessellation Model}\label{sec:poisson-process-model}

In this case study, we use the CTF to conduct statistical metamorphic testing (SMT) on a Poisson Line Tessellation (PLT) model. This model is of particular significance as it formed the case study of the paper that introduced the concept of SMT \cite{guderlei2007statistical}. As such it provides an ideal basis upon which to compare and contrast our CI-led approach against the conventional SMT approach. In particular, we show how the CTF can infer the same metamorphic test outcomes as the traditional SMT approach but from significantly fewer model executions. The code for this case study can be found in our open source repository\footnote{\url{https://github.com/CITCOM-project/CausalTestingFramework/tree/683e6c55/examples/poisson-line-process}}.

\begin{figure}[ht!]
    \centering
    \resizebox{0.5\textwidth}{!}{\input{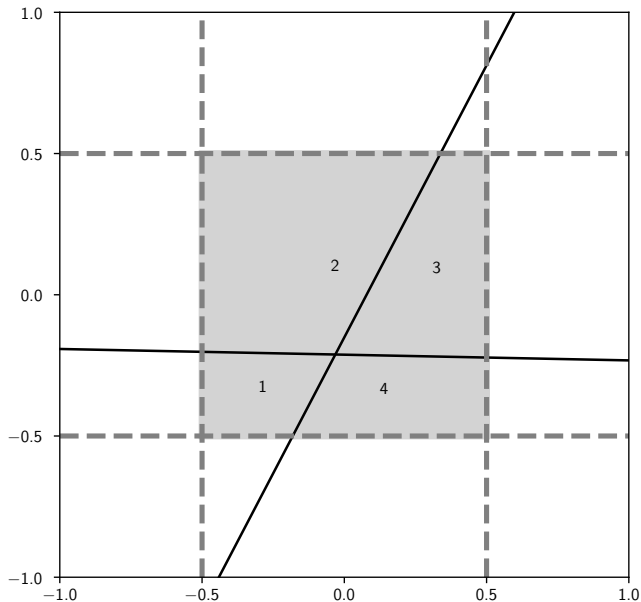}}
    \caption{A tessellation generated by the PLT model with a width ($W$), height ($H$), and intensity ($I$) of 1. There are two lines which intersect the sampling window ($L_t$, highlighted in grey). The intersection of these lines forms four polygons in total ($P_t$).}
    \label{fig:tessellation}
\end{figure}

\subsubsection{Subject System}
The PLT model uses a Poisson process to generate a series of lines that are positioned and oriented at random within a given sampling window to form a tessellation. While the behaviour of this model is predominantly random by design, it can be configured using three numerical input parameters to produce tessellations with predictable properties. In order to test these properties, we extract four numerical outputs from the resulting tessellation.


We selected this model because it has been the subject of prior research on SMT \cite{guderlei2007statistical} and has a number of well-characterised input-output relationships. In addition, Poisson process models are commonly used to model random processes for a range of important applications, including simulating road networks \cite{chetlur2018, morlot2012} and modelling photon arrival in 3D imaging \cite{shin2013parametric}. It is the stochastic yet predictable behaviour of Poisson process models that make them an interesting but difficult subject to test.

We now describe the behaviour of the PLT model, referring to the example tessellation in \Cref{fig:tessellation}.
The PLT model has three positive floating point input parameters: the width $W$ and height $H$ of a sampling window (shaded in grey in \Cref{fig:tessellation}), and the intensity $I$ of the Poisson process. Informally, the intensity parameter controls the average rate at which lines are placed. Given these inputs, the model generates a set of straight lines that intersect the origin-centred sampling window by drawing from a Poisson process on $[0, \infty) \times [0, 2\pi)$\footnotemark, where the orientation is uniformly distributed on $[0,\pi]$. The model then outputs the total number of lines intersecting the sampling window, $L_t$, and the number of polygons formed by the intersecting lines, $P_t$.

\footnotetext{The interval $[0,\infty)$ corresponds to the random distance of the lines to the origin, and the interval $[0, 2\pi)$ corresponds to the random angle of the point on the line that is closest to the origin. In the case of the orientation distribution, the upper interval bound is $\pi$ since rotating a line by an angle of $\pi$ (i.e. 180 degrees) leads to the same orientation.}

In \Cref{fig:tessellation}, for example, the inputs $W=H=I=1$ produce a tessellation in which there are two lines intersecting the sampling window ($L_t=2$) that form four polygons ($P_t=4$). Then, by dividing $L_t$ and $P_t$ by the sampling window area (i.e. $W\times H$), we obtain two further outputs corresponding to the number of lines and polygons per unit area ($L_u$ and $P_u$, respectively). Since $W=H=1$ in \Cref{fig:tessellation}, it follows that $L_u = L_t = 2$ and $P_u = P_t = 4$.

\subsubsection{Testing Activity}
In this case study, we replicate the SMT approach followed by Guderlei et al. in their seminal SMT paper \cite{guderlei2007statistical} to explore whether the CTF can achieve comparable results to traditional SMT approaches. Here we investigate whether the CTF can do so without the need for a large number of model executions (as is usually the case with SMT) and with a practically feasible amount of input from the tester.



As in the original paper, we expect the following two metamorphic relations to hold for the PLT model:
\begin{enumerate}
    \item Doubling $I$ should cause $P_u$ to increase by a factor of 4.
    \item $P_u$ should be independent of $W$ and $H$.
\end{enumerate}

\subsubsection{Data Generation}
We generated two sets of execution data. First, to obtain a ``gold standard'', we replicate the SMT approach followed in the original study \cite{guderlei2007statistical}. Specifically, we sampled 50 input configurations, with the bounds for width and height incremented together over the interval $\{n \in \mathbb{N} \mid 1 \leq n \leq 10\}$ (i.e. $W=H=1, W=H=2, \ldots, W=H=10$), such that the sampling window is always square, and the control and treatment values for intensity are powers of 2 up to 16. We then executed each configuration $100$ times to account for non-determinism, resulting in $5000$ model runs.

Second, to explore how the CTF enables us to re-use past execution data to infer the outcome of metamorphic test cases, we simulated an observational data set comprising 1000 executions of the PLT model. To produce this data set, we generated 1000 random input configurations using Latin hypercube sampling \cite{deutsch2012latin,moza2020lhsmdu} over the distributions $W, H \sim \mathbf{U}(0, 10)$ and $I \sim \mathbf{U}(0, 16)$. This sampling method provides even coverage of the input space and thus reduces our dependence on a statistical model to fill gaps in the data.

\newcommand{\risk}[4]{\frac{\expe{[#1 \mid #2=#4]}}{\expe{[#1 \mid #2=#3]}}}
\subsubsection{Causal Testing}
\begin{figure}[ht!]
  \centering
  \begin{tikzpicture}[every node/.style={circle,draw,inner sep=0pt}, scale=1]
    \node[vertex, minimum size=0.5cm] (W) at (0, 0.6) {$W$};
    \node[vertex, minimum size=0.5cm] (I) at (0, 0) {$I$};
    \node[vertex, minimum size=0.5cm] (H) at (0, -0.6) {$H$};
    \node[vertex, minimum size=0.5cm] (Lt) at (2, 0) {$L_t$};
    \node[vertex, minimum size=0.5cm] (Pt) at (4, 0) {$P_t$};
    \node[vertex, minimum size=0.5cm] (Lu) at (3, 0.6) {$L_u$};
    \node[vertex, minimum size=0.5cm] (Pu) at (6, 0) {$P_u$};
    \draw[edge] (I) to (Lt);
    \draw[edge] (Lt) to (Pt);
    \draw[edge] (Pt) to (Pu);
    \draw[edge] (W) to (Lt);
    \draw[edge] (H) to (Lt);
    \draw[edge] (W.15) ..controls (3.4, 1.4) and (3.5, 0.7) .. (Pt);
    \draw[edge, bend right=20] (H.345) ..controls (2, -0.6) and (3, -0.6) .. (Pt);
    \draw[edge, bend right=30] (H.east) ..controls (1, -0.5) and (2.6, -0.5) .. (Lu.south);
    \draw[edge] (W) to (Lu);
    \draw[edge] (Lt) to (Lu);
    \draw[edge] (W.30) ..controls (3.5, 1.7) and (5.5, 0.3) .. (Pu);
    \draw[edge] (H.330) ..controls (3, -0.8) and (3.8, -0.5) .. (Pu);
  \end{tikzpicture}
  \caption{A causal DAG for the PLT model.}
  \label{fig:pp-dag}
\end{figure}
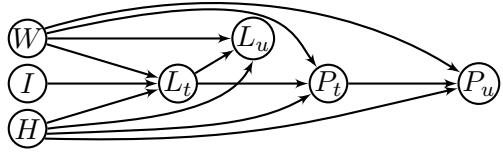

To begin causal testing, we specify our modelling scenario and causal DAG.
In line with the data generation process, our modelling scenario for this case study constrains the window to be a square with a maximum width (and height) of $10$ and places an upper limit of $16$ on the intensity parameter:
$$\{0 < W \leq 10, 0 < I \leq 16, W = H\}$$
We then construct the causal DAG shown in \Cref{fig:pp-dag} to model the following assumptions.
First, we add the causes of $L_t$ and $P_t$ based on the theoretical approximations $L_t \approx 2i(w+h)$ and $P_t \approx \pi i^2 wh$ \cite{chiu2013stochastic}. We do not, however, include a direct edge from $I$ to $P_t$ as the intensity ($I$) affects the number of polygons ($P_t$) indirectly through the number of intersecting lines ($L_t$). We then add the edge $L_t \to P_t$ since the number of polygons ($P_t$) is determined by the intersection of lines ($L_t$). Finally, we add edges from $W$ and $H$ to $L_u$ and $P_u$ since these quantities depend on the window area.

Having created our causal specification, we now perform a series of causal tests to investigate the two metamorphic relations mentioned above: (1) whether doubling $I$ causes $P_u$ to increase by a factor of 4, and (2) whether the sample window size has a causal effect on $P_u$.

\paragraph{Effect of $I$ on $P_u$}
First, we test whether doubling $I$ causes $P_u$ to increase by a factor of 4 for $I \in \{1, \ldots ,16\}$ and $W,H\in \{1, \ldots ,10\}$. Since we are interested in the multiplicative effect of $I$ on $P_u$, we use the \emph{risk ratio} (RR, see \Cref{sec:background}), which quantifies the factor by which the intervention (doubling $I$) causes the outcome change:
\begin{equation}\label{eqn:pp-rr}
    \text{RR} = \frac{\expe{[P_u \mid I=i_t]}}{\expe{[P_u \mid I=i_c]}}
\end{equation}
To estimate the RR using the CTF and observational data, we need to consider whether there is confounding bias in the data and design a regression model accordingly. To achieve this, we perform identification on the causal DAG shown in \Cref{fig:pp-dag}, revealing that there is no confounding over the effect of $I$ on $P_u$ in this scenario. Therefore, we do not need to include additional terms for confounders in our regression model. However, because we expect $P_u$ to vary quadratically with $I$, we opt to include the term $I^2$. This assumption is informed by domain expertise \cite{guderlei2007statistical} but can be validated by varying $I$ and observing changes to $P_u$. This process yields a regression model of the following form:
\begin{equation}\label{eqn:pp-reg}
    P_u \sim x_1 I + x_2 I^2
\end{equation}
We then apply the regression model to our observational data to obtain a causal estimate of the RR (\Cref{eqn:pp-rr}). 

\paragraph{Effect of $W$ on $P_u$}
Second, we test whether the sample window size $W$ has a causal effect on $P_u$. Since we are only interested in whether there is \emph{some} effect, we use the \emph{average treatment effect} (ATE, see \Cref{sec:background}), which quantifies the additive change in outcome caused by the intervention (increasing $W$):
\begin{equation}\label{eqn:pp-ate}
    \text{ATE} = \expe{[P_u \mid W=w_t]} - \expe{[P_u \mid W=w_c]}
\end{equation}

Ordinarily, to investigate whether $W$ affects $P_u$ using SMT, we would need to execute a fresh, customised set of test cases, this time fixing the value of $I$ and varying $W$. In the CTF, however, we can infer this effect from the \emph{same} 1000 model runs (i.e. re-using data from \emph{previous} test executions to predict \emph{new} test outcomes). 

To achieve this, we start by performing identification on the causal DAG (\Cref{fig:pp-dag}) for the effect of $W$ on $P_u$, once again revealing the absence of confounding. We then modify the regression model shown in \Cref{eqn:pp-ate} to include terms for $W$ and $W^{-1}$, reflecting the hypotheses that $W$ \emph{does} affect $P_u$ and that they share an inverse relationship (this can be validated by varying $W$ and observing $P_u$). Although $I$ is not a confounder here, we retain the $I$ and $I^2$ terms to increase the accuracy of the model. The DAG in \Cref{fig:pp-dag} allows us to show that this does not bias our predictions. This process yields the following regression model:
\begin{equation}\label{eqn:reg-pp-2}
    P_u \sim x_1 W + x_2 W^{-1} + x_2 I + x_2 I^2
\end{equation}

We then apply this model to the \emph{original} data to obtain a causal estimate for the ATE (\Cref{eqn:pp-ate}). The effect of $H$ could be investigated similarly, but we omit this due to space constraints.





\subsubsection{Results}
\begin{table}[ht!]
  \centering
  \caption{RR of doubling $I$ under different values of $W$ and $H$. The bottom row gives the value estimated using regression. Bold values round to 3, violating the expected behaviour.}
  \begin{small}
    \begin{tabular}{rrcccc}
      \toprule
      \multicolumn{1}{c}{$W$} & \multicolumn{1}{c}{$H$} & $\risk{P_u}{I}{1}{2}$ & $\risk{P_u}{I}{2}{4}$ & $\risk{P_u}{I}{4}{8}$ & $\risk{P_u}{I}{8}{16}$ \\
      \midrule
      1  & 1  & \textbf{2.5888} & \textbf{3.4461} & 3.6178 & 3.6187 \\
      2  & 2  & \textbf{3.0359} & 3.5410          & 3.6003 & 3.7264 \\
      3  & 3  & 3.5025          & 3.5945          & 4.0191 & 3.6545 \\
      4  & 4  & \textbf{3.1138} & 3.5285          & 4.1562 & 3.7290 \\
      5  & 5  & 3.6686          & 3.7686          & 3.9408 & 3.8751 \\
      6  & 6  & 3.6933          & 3.6988          & 3.9219 & 3.9707 \\
      7  & 7  & 3.7127          & 3.6271          & 3.9862 & 3.9370 \\
      8  & 8  & \textbf{3.4957} & 3.8300          & 3.8861 & 4.0110 \\
      9  & 9  & 3.5633          & 4.0009          & 3.9342 & 3.9338 \\
      10 & 10 & 3.8275          & 3.7525          & 4.0128 & 4.0181 \\
      \midrule
      \multicolumn{2}{l}{CTF Estimate} & \textbf{2.8280} & \textbf{3.1711} & \textbf{3.4772} & 3.6993 \\
      \bottomrule
    \end{tabular}
  \end{small}
  \label{table:smt-results}
\end{table}

\Cref{table:smt-results} shows the results for our investigation into the effect of $I$ on $P_u$ using \Cref{eqn:pp-reg}. The first 10 rows show the RRs obtained via the conventional SMT approach for various values of $W$ and $H$, and the final row shows the RRs estimated using the CTF and observational data. The discrepancy between the regression estimations and the SMT results are likely due to \Cref{eqn:pp-reg} not including $W$ and $H$ terms, which the SMT results explicitly control for. However, this does not represent a biased result as \Cref{fig:pp-dag} shows there is no confounding.

These results show that both approaches identify an inconsistency between the metamorphic relations and implementation from the original study \cite{guderlei2007statistical}: for lower values of $W$, $H$, and $I$, doubling $I$ causes $P_u$ to increase by a factor that is closer to three than four, meaning our metamorphic relation is not satisfied. This is a particularly interesting result since $P_u$ should be independent of $W$ and $H$. 

Furthermore, these results show that the CTF was able to identify the same discrepancy as conventional SMT, but using a fifth of the data. This highlights the potential of CI-led approaches to offer economical alternatives to testing techniques that depend on repeated potentially costly executions of the SUT.



\Cref{table:width-results} shows the results of our investigation into the effect of $W$ on $P_u$ using \Cref{eqn:reg-pp-2} and the same random 1000 data points as for the last row of \Cref{table:smt-results}. Here, each row shows how $P_u$ changes when $W$ is increased from $W_{c}$ to $W_{t}$ with the intensity fixed to $I=1$. According to the original study \cite{guderlei2007statistical}, changes to $W$ should \emph{not cause} a change to $P_u$. Our results show that this property holds for all but the first row because these rows have confidence intervals that contain zero, meaning there is no statistically significant causal effect. However, the 95\% confidence intervals for the first row of \Cref{table:width-results} show that, when $W$ is increased from $W=1$ to $W=2$, there is a statistically significant causal effect on $P_u$ of $-7.3786$. Although they are wide, indicating that the causal effect is variable, the fact that they do not contain zero indicates that the effect is statistically significant.

\begin{table}[h!]
  \centering
  \caption{Estimated ATE of increasing W from W$_\text{c}$ to W$_\text{t}$ on $P_u$ with $I=1$ in the PLT model with 95\% confidence intervals.}
  \label{table:width-results}
  \begin{tabular}{cccl}
  \toprule
  W$_\text{c}$ &  W$_\text{t}$ &     ATE &  95\% CIs \\
  \midrule
       1 &          2 & -7.3786 & [-13.9182,  -0.8390] \\
       2 &          3 & -2.7097 & [ -9.8029,   4.3836] \\
       3 &          4 & -1.5424 & [-11.1209,   8.0361] \\
       4 &          5 & -1.0755 & [-13.7084,  11.5574] \\
       5 &          6 & -0.8421 & [-16.7413,  15.0572] \\
       6 &          7 & -0.7087 & [-19.9729,  18.5556] \\
       7 &          8 & -0.6253 & [-23.3084,  22.0578] \\
       8 &          9 & -0.5697 & [-26.7043,  25.5649] \\
       9 &         10 & -0.5308 & [-30.1383,  29.0767] \\
  \bottomrule
  \end{tabular}
\end{table}

This conflicting result indicates a problem with either the program or the metamorphic property. In this case, we believe that the problem stems from basic geometry: lines are less likely to intersect a smaller sample window. As the sample window becomes larger, there is more area to average over so $P_u$ becomes more reliable. Therefore, the metamorphic relations should ideally specify a minimum window size to which they apply.

Overall, this case study has provided evidence related to all three research questions.

\paragraph{\textbf{RQ1}}
In this case study, we demonstrated the CTF's ability to reproduce published SMT results from \cite{guderlei2007statistical} using a sample of randomly generated test data. First, we estimated the risk ratio of doubling $I$. \Cref{table:smt-results} shows that our regression model was able to give sufficiently accurate results to discover an inconsistency that was also revealed by SMT, even though it did not explicitly control for $W$ and $H$ like SMT did. In the second part of the case study, we investigated this inconsistency further, and estimated the ATE of increasing $W$ on $P_u$. While we expected this to be zero, \Cref{table:width-results} shows that there is actually a statistically significant negative relationship when we increase $W$ from 1 to 2.


\paragraph{\textbf{RQ2}}
This case study demonstrated the CTF's ability to find the same bugs as SMT using only a fraction of the data. Furthermore, the second part of this case study involved using the \emph{same data} as for the first part to test a \emph{different relationship} after having discovered a potential bug in the system. By contrast, the traditional SMT approach would need to perform additional controlled runs of the system, which vary $W$ while holding $I$ constant, to test this new property.

\paragraph{\textbf{RQ3}}
The DAG for this case study, shown in \Cref{fig:pp-dag}, required minimal effort to construct.
There are no internal variables here, and the relationship between the inputs and outputs is well documented in \cite{guderlei2007statistical}.
The main drawback is the requirement for the domain expert to have an approximate idea of the ``shape'' of the relationships between different variables, for example that $P_u$ varies with $I^2$ rather than just $I$, in order to obtain accurate estimates.\\

This case study has shown that not only can we conduct SMT using the CTF, but we can do so \emph{using previous execution data} and \emph{less data} than a traditional SMT method. Furthermore, we demonstrated how this approach allowed us to refine our metamorphic relations and find faults \emph{without running the SUT additional times}.



\subsection{Cardiac Action Potential Model}
\label{sec:cardiac-ap-model}
In this case study, we use the  CTF to conduct sensitivity analysis on the Luo-Rudy 1991 ventricular cardiac action potential model \cite{luo1991model} (LR91) in a straightforward and efficient way. Sensitivity analysis is commonly used to validate and verify scientific models, with a specific focus on identifying which inputs have the greatest impact on model outputs \cite{kleijnen1995verification, sarrazin2016global}. Here, we take a CI-led approach and measure the ATE of several input parameters on one output, quantifying the extent to which this output is affected by changes to the inputs. As test oracles, we construct a series of metamorphic relations that capture the expected magnitude and direction of each ATE.

Throughout this case study, we follow part of an existing study \cite{chang2015bayesian} that conducts uncertainty and sensitivity analysis on LR91 using a Gaussian Process Emulator (GPE) \cite{rasmussen2003gaussian} trained on runs of the model. This work provides an invaluable source of domain expertise that precisely quantifies several cause-effect relationships between the inputs and outputs of LR91 that we use as the basis for constructing our metamorphic relations. However, in contrast to the data-driven approach employed in the original study, we employ causal knowledge and domain expertise to justify and hand-craft a simple regression model that reaches the same conclusions. The code for reproducing this case study can be found in our open source repository\footnote{\url{https://github.com/CITCOM-project/CausalTestingFramework/tree/683e6c55/examples/lr91}}.


\subsubsection{Subject System}
The Luo-Rudy 1991 ventricular cardiac action potential model \cite{luo1991model} (LR91) is a mathematical model comprising a system of differential equations that describe the rapid rise and fall in the voltage across the membrane of a mammalian ventricular cell. This characteristic rise and fall in voltage is referred to as an \emph{action potential}. The behaviour of this model is controlled by 24 constants, 8 rate variables, 8 state variables, and 25 algebraic variables.

We selected LR91 as a case study as it follows a different modelling paradigm to our other subject systems and has supported extensive and important research into cardiovascular physiology. Furthermore, amongst its vast and largely uncertain input space, LR91 has several well-characterised input-output relationships suitable for causal analysis.

\begin{figure}[!h]
    \centering
    \includegraphics[width=.5\linewidth]{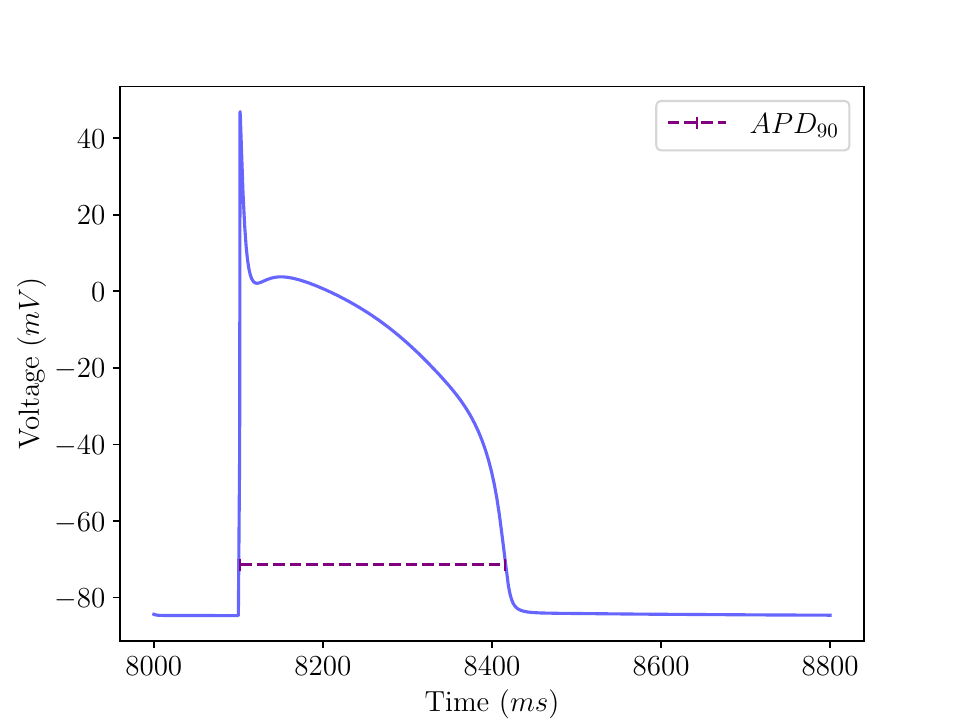}
    \caption{An example action potential produced by the Luo-Rudy 1991 model, simulating the rise and fall of voltage across a mammalian ventricular cell, and the output of interest: $APD_{90}$.}
    \label{fig:lr91-output}
\end{figure}

An example action potential produced by LR91 is shown in \Cref{fig:lr91-output}, demonstrating the rapid rise (known as depolarisation) and corresponding fall (repolarisation) of the voltage over time. In this case study, we quantify the effect of six conductance-related input parameters on one attribute of the action potential: \emph{action potential duration to 90\% of repolarisation} ($APD_{90}$). That is, the amount of time taken for the action potential to repolarise by 90\%. This output concerns the falling phase of the action potential in which the cell returns to its resting voltage \cite{grider2019physiology} and is shown in \Cref{fig:lr91-output}.


\subsubsection{Testing Activity}
In this case study, we replicate part of an existing study \cite{chang2015bayesian} that conducts a sensitivity analysis on LR91 using a Gaussian Process Emulator (GPE) \cite{rasmussen2003gaussian}. In short, the approach in \cite{chang2015bayesian} trained a GPE on 200 runs of LR91, with input configurations sampled via Latin Hyper Cube Sampling \cite{stein1987large} from a series of normalised uniform design distributions to ensure even coverage of the input space.
The GPE was then used to calculate the expectation of a given output, conditional on an input of interest, to quantify the effect of varying each of the six inputs on the eight output parameters, over the range of the design distribution.

From a CI perspective, we can obtain similar information by computing the ATE of each input on each output over the range of the design distribution. Specifically, we can set our control value to the mean value of the design distribution and uniformly increment our treatment value from the minimum to the maximum value of the design distribution. This yields a series of ATEs that quantify the expected change in output caused by changing the input parameters by specific amounts above and below their mean, revealing the magnitude of each input's effect on the outputs.

Due to space limitations, we limit our analysis to the effect of the six inputs on one output, $APD_{90}$. We have selected this output because the original paper uses it to illustrate the approach. Based on the results reported in \cite{chang2015bayesian}, we expect the following metamorphic properties to hold:
\begin{enumerate}
    \item Increasing the parameters $G_K$, $G_b$, and $G_{K1}$ should cause $APD_{90}$ to decrease.
    \item Increasing the parameter $G_{si}$ should cause $APD_{90}$ to increase.
    \item Increasing the parameters $G_{Na}$ and $G_{Kp}$ should have no significant effect on $APD_{90}$.
    \item The following monotonic relationship should hold over the (absolute) magnitude of the parameters' effects:
    $$\lvert APD_{90}^{G_{si}} \rvert > \lvert APD_{90}^{G_K} \rvert > \lvert APD_{90}^{G_b} \rvert > \lvert APD_{90}^{G_{K1}} \rvert$$
\end{enumerate}


\subsubsection{Data Generation}
To gather data from LR91, we followed the same approach as \cite{chang2015bayesian}, where the 200 input configurations were sampled from the design distributions using Latin Hyper Cube sampling and then normalised. We then executed each of these input configurations on an auto-generated Python implementation of LR91 from the cellML modelling library \cite{lr912022cellml}. We extended this implementation to enable us to sample the input values via Latin Hyper Cube sampling and automatically extract the outputs\footnote{Our LR91 model is available at: \url{https://github.com/AndrewC19/LR91/tree/769e7ff}}.

\subsubsection{Causal Testing}
To approach sensitivity analysis as a CI problem, we first specify our modelling scenario and causal DAG. For this set of tests, the modelling scenario constrains each input to the range of its uniform design distribution (as specified in the original paper \cite{chang2015bayesian}):
\begin{align*}
    \{&17.250 \leq G_{Na} \leq 28.750, 0.0675 \leq G_{si} \leq 0.1125, 0.2115 \leq G_K \leq 0.3525,\\
    &0.4535 \leq G_{K1} \leq 0.7559, 0.0137 \leq G_{Kp} \leq 0.0229, 0.0294 \leq G_b \leq 0.0490\}
\end{align*}
As in the original study, these input values were then normalised to the range [0, 1].

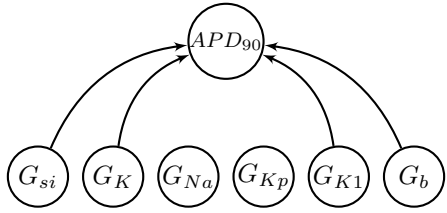
\begin{figure}[!t]
    \centering
    \begin{tikzpicture}[every node/.style={circle,draw,inner sep=0pt}, scale=1]
        \node[vertex, minimum size=0.8cm] (si) at (0, 0) {\small $G_{si}$};
        \node[vertex, minimum size=0.8cm] (K) at (1, 0) {\small $G_K$};
        \node[vertex, minimum size=0.8cm] (Na) at (2, 0) {\small $G_{Na}$};
        \node[vertex, minimum size=0.8cm] (Kp) at (3, 0) {\small $G_{Kp}$};
        \node[vertex, minimum size=0.8cm] (K1) at (4, 0) {\small $G_{K1}$};
        \node[vertex, minimum size=0.8cm] (b) at (5, 0) {\small $G_b$};
        \node[vertex, minimum size=0.8cm] (APD90) at (2.5, 1.8) {\scriptsize $APD_{90}$};
        \draw[edge] (si) to [bend left] (APD90) ;
        \draw[edge] (K) to [bend left] (APD90);
        \draw[edge] (K1) to [bend right] (APD90);
        \draw[edge] (b) to [bend right] (APD90);
    \end{tikzpicture}
    \caption{LR91 modelling scenario's Causal DAG, where the sensitivity of $APD_{90}$ to each conductance input is computed as the causal effect (ATE).}
    \label{fig:LR91-dag}
\end{figure}

We then specify the expected cause-effect relationships (and absence thereof) as the causal DAG shown in \Cref{fig:LR91-dag}. While not essential, we include the isolated nodes $G_{Na}$ and $G_{Kp}$ in our DAG to make our expectation for the absence of a causal effect explicitly clear. For each relationship, we then create a suite of causal test cases covering a series of interventions that incrementally increase/decrease the value of the inputs over the range of the design distribution. For each input, this is achieved by setting the control value to 0.5 (the mean) and uniformly sampling 10 treatment values over the range [0, 1]. This produces a total of 10 test cases per input that vary its value from 0.5 to each of the treatment values: $[0,\,0.1,\,0.2,\,...\,1.0]$. Using the CTF, we then perform identification and estimation. Here, the cause-effect relationships are straightforward and there is no confounding to adjust for, enabling us to fit a regression model $APD_{90} \sim x_0 + x_1 G_z$ for each input $z \in \{ si, K, Na, Kp, K1, b\}$. Using these models, we then predict the ATE and 95\% confidence intervals for each test.

\subsubsection{Results}
\begin{figure}[!ht]
    \centering
    \includegraphics[width=.8\linewidth]{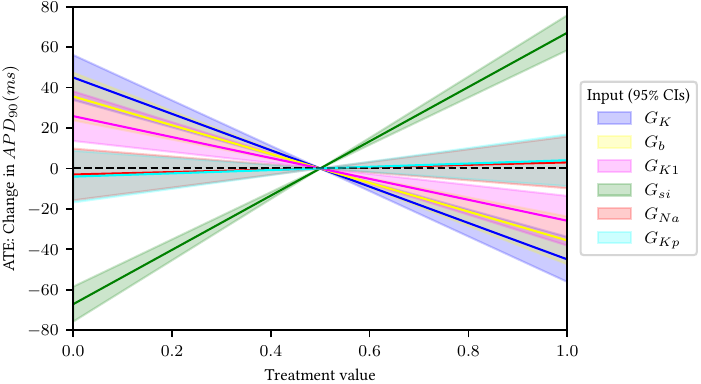}
    \caption{Sensitivity of $APD_{90}$ in response to changes to the mean value of input parameters in LR91.}
    \label{fig:LR91-results}
\end{figure}

The results, as summarised in \Cref{fig:LR91-results}, show that all expected metamorphic relationships pass with statistical significance (95\% confidence intervals do not contain 0) and are visually similar to Figure 5 in the original study \cite{chang2015bayesian}. Specifically, the first metamorphic relation holds as $G_K$, $G_{K1}$, $G_b$ have negative effects, the second metamorphic relationship holds because $G_{si}$ has a positive effect, and the third metamorphic relation holds as 
$G_{Na}$ and $G_{Kp}$ have no significant effect. Furthermore, the fourth metamorphic relation holds as the gradients corresponding to these effects reveal that the effect sizes follow the expected monotonic relationship: $\lvert APD_{90}^{G_{si}} \rvert > \lvert APD_{90}^{G_K} \rvert > \lvert APD_{90}^{G_b} \rvert > \lvert APD_{90}^{G_{K1}} \rvert$.

This case study has provided insights into \textbf{RQ1} and \textbf{RQ3}. As a result of following the data generation approach of the original paper, however, this case study did not afford us the opportunity to evaluate the efficiency of the CTF.

\paragraph{\textbf{RQ1} Accuracy} 
In this case study, we used the CTF to conduct a sensitivity analysis on the LR91 model, achieving visually similar results to an existing approach that employed a GPE \cite{chang2015bayesian}. However, we achieved this using a significantly simpler statistical model whose design was informed by causal reasoning as opposed to associations within the data. This contrast between a model-based and black-box approach to reasoning about system behaviour raises an interesting discussion around \emph{explainability} that we return to in \Cref{sec:discussion}.

\paragraph{\textbf{RQ3} Practicality}
In this case study, the process of specifying the causal DAG was straightforward and required minimal domain expertise that were easily gleaned from the original study \cite{chang2015bayesian}. Since the resulting DAG contained no confounding (\Cref{fig:LR91-dag}), the regression model for each causal test simply regressed the input-under-test against against $APD_{90}$. By contrast, Gaussian Processes (as used in the original study) have several practical limitations, including the need to specify an appropriate kernel function for the problem at hand \cite{nevin2021physics}, and a complexity of $O(n^3)$ that hinders the feasibility of the approach when dealing with large amounts of data \cite{rasmussen2006gaussian}.



Overall, in this case study, we have shown that the CTF reaches the same conclusions as the original study. However, the CTF achieves this by using a simpler, more practical statistical model guided by causality instead of associations within the data.
\subsection{Covasim: Experimental Casual Testing}
\label{sec:covasim-exp}

In this case study, we demonstrate the ability of the CTF to conduct statistical metamorphic testing (SMT) of Covasim  \cite{kerr2021covasim} using the experimental mode of the CTF (\Cref{sec:data-collection}). That is, isolating the causal effect of interest via strategic executions of the SUT, rather than applying graphical CI to observational data. Our aim here is to provide evidence to support our claim that metamorphic testing is a fundamentally causal activity that can be framed and solved as a problem of CI. The code for this case study can be found in our open source repository\footnote{\url{https://github.com/CITCOM-project/CausalTestingFramework/tree/683e6c55/examples/covasim_/vaccinating_elderly}}.

\subsubsection{Subject System}
Covasim is the epidemiological agent-based model that was introduced as a motivating example in \Cref{sec:background}. As a brief reminder, it is a complex, real-world scientific model that is primarily used to simulate detailed COVID-19 scenarios in order to evaluate the impact of various interventions, such as vaccination and contact tracing \cite{kerr2021covasim}, in specific demographics. These scenarios are configured via 64 input parameters and described by 56 time-series outputs. It has been used to inform a number of important policy decisions across a range of countries, including the UK, US, and Australia \cite{panovska2020potential, kerr2021controlling, panovska2020determining, stuart2020role},



We cover two testing scenarios using Covasim. In this section, we elaborate upon our example from \Cref{sec:background} and use the experimental mode of the CTF to test the effect of prioritising the vaccination of elderly people on several vaccine-related outcomes, revealing an interesting bug in the process. Then, in \Cref{sec:covasim-obs} we test the effect of increasing the $\beta$ parameter (transmissibility) on cumulative infections using execution data from other tests (i.e. data that has not been customised to explore this specific effect).

\subsubsection{Testing Activity}
Revisiting our example from \Cref{sec:causal-testing-framework}, our aim is to determine the effect of prioritising vaccination for the elderly on the following outputs: cumulative infections, number of doses given, maximum number of doses per agent, and number of agents vaccinated.

Our expectation here is that prioritising the elderly should lead to an increase in infections. This is because we are less likely to vaccinate agents in the model with a greater propensity for spreading the virus (e.g. younger individuals who attend a school or workplace). We also expect the number of vaccines and doses administered to decrease as there are fewer elderly agents in the model. In contrast, the maximum number of doses should not change, as the vaccine is set to be administered at most two times per agent.

\subsubsection{Data Generation}
We executed the model under two input configurations 30 times each using an experimental data collector (see \Cref{sec:data-collection}) for every test. For both input configurations, we used the default Covasim parameters, but fixed the simulation length to 50 days, initial infected agents to 1000, population size to 50,000, and made the default Pfizer vaccine available from day seven. However, for the second configuration, we also sub-targeted (prioritised) vaccination to the elderly using the \texttt{vaccinate\_by\_age} method from the Covasim vaccination tutorial\footnote{\url{https://github.com/InstituteforDiseaseModeling/covasim/blob/master/examples/t05_vaccine_subtargeting.py}}.

\subsubsection{Causal Testing}
Although we provide a causal DAG (\Cref{ex:causal-spec}) as an illustrative example for this scenario in \Cref{sec:causal-testing-framework}, it is not necessary to perform identification since, under the experimental mode of operation (\Cref{sec:data-collection}), we explicitly control for potential biases. Consequently, there is no confounding to adjust for in the resulting data, enabling us to calculate the ATE directly by contrasting the average cumulative infections produced by the control (vaccinate everyone) and treatment executions (prioritise the elderly).

\subsubsection{Results}
As expected, prioritising the elderly causes the cumulative infections to increase (ATE: 2399.7, 95\% CIs: [2323.7, 2475.8]) and causes no change to the maximum doses (ATE: $8.9\mathrm{\times10}^{-16}$, 95\% CIs: [$3.7\mathrm{\times10}^{-17}$, $4.1\mathrm{\times10}^{-16}$]).

However, when we examine the number of doses given (which, as stated in \Cref{ex:causal-spec}, we would expect to remain fixed), the tests in fact show that the SUT erroneously causes the number of doses administered and the number of people vaccinated to increase sharply by 481351 (95\% CIs: [480550, 482152]) and 483506 (95\% CIs: [482646, 484367]), respectively. This is an obvious and potentially problematic bug, as it reveals that more agents have been vaccinated than there are agents in the simulation (by a factor of 9.7).

We raised an issue\footnote{\url{https://github.com/InstituteforDiseaseModeling/covasim/issues/370}} on Covasim's GitHub repository to report this bug in September 2021 and the Covasim developers replied in November confirming that the bug had been fixed for version 3.1. Although the developers did not explain the cause of the bug nor how it was fixed, the change log for version 3.1 stated the following: \emph{Rescaling now does not reset vaccination status; previously, dynamic rescaling erased it.}



This testing scenario has provided insights related to \textbf{RQ2} and \textbf{RQ3}. Due to employing the experimental mode of the CTF (\Cref{sec:data-collection}), we have not inferred test outcomes from observational data and therefore this case study does not offer any insights into the accuracy associated with the observational approach.

\paragraph{\textbf{RQ2 (Efficiency)}}
We used the experimental mode of the CTF to quantify the effect of introducing a vaccination policy on a number of variables, essentially conducting SMT in the conventional way. We repeated both the source and follow-up test cases for each metamorphic relation 30 times for each test (of which there were four), requiring a total of $30 \times 2 \times 4 = 240$ executions of Covasim. We show how, under the experimental mode of operation, the CTF can conduct SMT in the conventional way and demonstrate that, in situations where observational data is unavailable, the CTF can match the efficiency of conventional SMT. 

\paragraph{\textbf{RQ3 (Effort)}}
The amount of human effort required to apply the CTF was low. We did not need to provide a DAG and we did not need to specify a regression model. Instead, the main expenditure of human effort in this case study lies in the process of implementing the test harness for experimental data collection; a step that is required for most model-based testing techniques.

Overall, this case study has demonstrated how the CTF can also be employed under the experimental mode of operation to essentially conduct a conventional SMT approach. This revealed a problematic bug related to vaccination, highlighting the importance of applying metamorphic testing in the scientific context.

\subsection{Covasim: Observational Causal Testing}
\label{sec:covasim-obs}
We now consider the effect of increasing transmissibility ($\beta$) on cumulative infections, but this time applying the CTF to simulated confounded observational data. Here we compare the outcomes inferred by the CTF to the same outcomes achieved using a conventional SMT approach. Our goal here is to understand whether the CTF can operate accurately and efficiently within the challenging context presented by Covasim.

This case study presents a significant testing challenge. There are 156 distinct locations that can be simulated in Covasim that will lead to differing rates of transmission.  This is because different locations are modelled with different age distributions and household contact patterns, leading to differences in key attributes of the population, such as susceptibility, that also affect infection dynamics.

Furthermore, Covasim is non-deterministic. Each metamorphic test requires multiple repeats of the source and follow-up tests, making conventional SMT extremely costly in this context. For example, if we repeat both the source and follow-up test cases 30 times for each location, we would need to run $30 \times 2 \times 156 = 9360$ simulations. Although we do not provide precise timing measurements, on a moderate specification machine\footnote{\label{machine-spec}MacBook Pro, Core i7, 16GB 2133 MHz LPDDR3 RAM} each of these runs takes between 1 and 2 minutes to complete, requiring between 156 and 312 hours to run all simulations (without parallelisation). The code for this case study can be found in our open source repository\footnote{\url{https://github.com/AndrewC19/covasim_case_study/tree/65bc40a}}.



\subsubsection{Data Generation}
When reasoning about transmissibility and the spread of COVID-19 using Covasim, there are several parameters that can affect the output. These include the variant of the virus and population characteristics such as age and household size, with older populations being more susceptible to infection and higher household contacts leading to quicker viral spread. These population characteristics cannot be specified directly, but can be indirectly altered by selecting a geographical location.

For this case study, we generate two sets of data.
First, we directly apply a conventional SMT approach to Covasim in which we execute the model 30 times with $\beta=0.016$ and $\beta=0.02672$ for each location, before averaging and contrasting their respective cumulative infections. We select these values of $\beta$ as they correspond to the $\beta$ values for the Beta and Alpha variants of COVID-19 available in Covasim.

Second, we simulate (uncontrolled) observational data.
To achieve this, we assign a different dominant variant (Alpha, Beta, Delta, Gamma) to each location at random, each of which has its own specific $\beta$ value ($\beta_\alpha = 0.02672, \beta_\beta = 0.016, \beta_\delta = 0.0352, \beta_\gamma = 0.0328)$. For each location, we then create a normal distribution centred around the location-specific $\beta$ value and a standard deviation of $0.002$. We select this standard deviation to give some variance in the exact value of $\beta$ used for each run of the location, without introducing too much overlap with other variants. We then run 30 simulations for each location, sampling a fresh $\beta$ value from its distribution on each run. For all simulations, we use a population size of 1 million individuals, 1000 initially infectious individuals, and a duration of 200 days. This results in a data set comprising 4680 simulations (30 per location).

\subsubsection{Causal Testing}
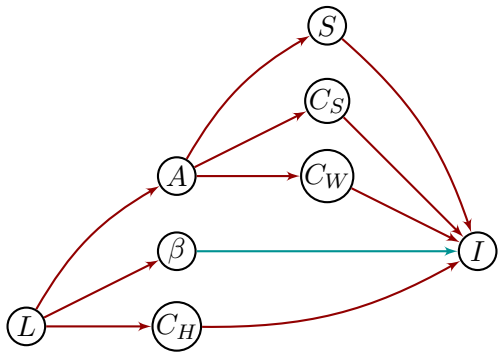
\begin{figure}[h!]
  \centering
  \begin{tikzpicture}[every node/.style={circle,draw, inner sep=0}, scale=1]
    \node[vertex, minimum size=0.5cm] (L) at (0, -2) {$L$};
    \node[vertex, minimum size=0.5cm] (B) at (2, -1) {$\beta$};
    \node[vertex, minimum size=0.5cm] (I) at (6, -1) {$I$};
    \node[vertex, minimum size=0.5cm] (A) at (2, 0) {$A$};
    \node[vertex, minimum size=0.5cm] (CH) at (2, -2) {$C_H$};
    \node[vertex, minimum size=0.5cm] (S) at (4, 2) {$S$};
    \node[vertex, minimum size=0.5cm] (CS) at (4, 1) {$C_S$};
    \node[vertex, minimum size=0.5cm] (CW) at (4, 0) {$C_W$};
    \draw[edge, color=causal]      (B) to (I);
    \draw[edge, color=association] (L) to (B);
    \draw[edge, color=association] (L) to [bend left=15] (A);
    \draw[edge, color=association] (L) to (CH);
    \draw[edge, color=association] (A) to [bend left=15] (S);
    \draw[edge, color=association] (A) to (CS);
    \draw[edge, color=association] (A) to (CW);
    \draw[edge, color=association] (S) to [bend left=15] (I);
    \draw[edge, color=association] (CS) to (I);
    \draw[edge, color=association] (CW) to (I);
    \draw[edge, color=association] (CH) to [bend right=15] (I);
  \end{tikzpicture}
  \caption{A causal DAG for the Covasim modelling scenario where the \textcolor{causal}{causal effect} of $\beta$ on $I$ is \textcolor{association}{confounded}. Here, $L$ denotes the location; $A$ denotes the average age of the population; $\beta$ denotes the transmissibility of the virus; $C_H$, $C_S$, $C_W$ denote household, school, and workplace contacts; $S$ denotes average susceptibility of the population; and $I$ denotes the total cumulative infections.}
  \label{fig:covasim-dag}
\end{figure}

To begin causal testing, we form our causal specification by specifying a modelling scenario and the causal DAG shown in \Cref{fig:covasim-dag}. Our modelling scenario uses the default Covasim parameters apart from $\beta$ (the input-under-study) and the location. We also fixed the duration, population size, and initial infected agents as follows:$$\{\mathtt{days}=200,\ \mathtt{pop\_size}=1000000,\ \mathtt{pop\_infected}=1000\}$$

Next, we consider the adjustment sets implied by the causal DAG in \Cref{fig:covasim-dag}. While there are many possible adjustment sets for this causal DAG, there are three notable choices to discuss. 

First, we could use the smallest adjustment set $\{L\}$. This has the advantage of conditioning on the least variables, but restricts estimation to using location-specific data only (i.e. not borrowing data from \emph{similar} locations). Second, we could use $\{A, C_H\}$. This would enable us to additionally borrow information from locations that have similar average ages and household contacts. From an information theoretic standpoint, however, this is not a sensible choice as the average age is not a good measure for the shape of the age distribution (two populations with a similar average age may have vastly different age distributions). To this end, we can consider a third adjustment set $\{S, C_S, C_W, C_H\}$. Here, we replace $A$ with the variables related to age that directly affect cumulative infections: the number of school and workplace contacts (assignment to these environments is determined by age) and susceptibility (which varies with age).

For this case study, we select this third adjustment set on the basis that it most accurately captures the key causal measures while allowing us to borrow data from other locations that are similar with respect to these attributes.
This yields the following closed-form statistical expression that is capable of directly estimating the causal effect (CATE) of interest:
$$CATE = \E[I \mid \beta = 0.02672, S, C_S, C_W, C_H] - \E[I \mid \beta = 0.016, S, C_S, C_W, C_H]$$
Then, to estimate the value of this estimand, we implement a regression model of the following form, where $Z$ is our adjustment set $\{S, C_S, C_W, C_H\}$ and each variable in this adjustment set has three coefficients: $x^z_1, x^z_2, x^z_3$:
$$I \sim x_0 + x_1 ln(\beta) + \sum_{z \in Z} x^z_1 ln(z) + x^z_2 ln(z)^2 + x^z_3 ln(z) ln(\beta)$$
This regression model encodes three key assumptions. First, due to the exponential nature of viral infection, we apply a log transformation to the variables on the right-hand-side of the equation \cite{stock2003introduction, benoit2011linear}.
Second, we add a quadratic term for each of our adjusted variables. This captures the possibility of curvilinear relationships between $I$ and the parameters. 
Third, we include an interaction term between $\beta$ and each of our adjusted parameters. This captures our expectation that the effect of $\beta$ on cumulative infections is moderated by the number of contacts and susceptibility of the population, and enables the model to make location-specific estimates i.e. conditional ATEs (CATEs; see \Cref{sec:causal-inference})\footnote{We formed these assumptions by varying each parameter in isolation and observing the change in cumulative infections. An epidemiologist, however, may know more precise characterisations of these relationships a priori.}.

At this point, we have specified a causally-valid statistical model that is capable of directly estimating the causal effect of $\beta$ on cumulative infections for each location separately. We can therefore compute the average values for the variables $S$, $C_S$, $C_W$, and $C_H$ for each location using our observational data, and substitute these into the model alongside the values $\beta = ln(0.016)$ and $\beta = ln(0.2672)$\footnote{We take logarithms of the treatment and control values here to maintain the interpretability of our estimate.}. By contrasting the respective estimates for $I$, we obtain an estimate of the causal effect for each location in Covasim.

\subsubsection{Results}
\begin{figure}
    \centering
    \includegraphics[width=\linewidth]{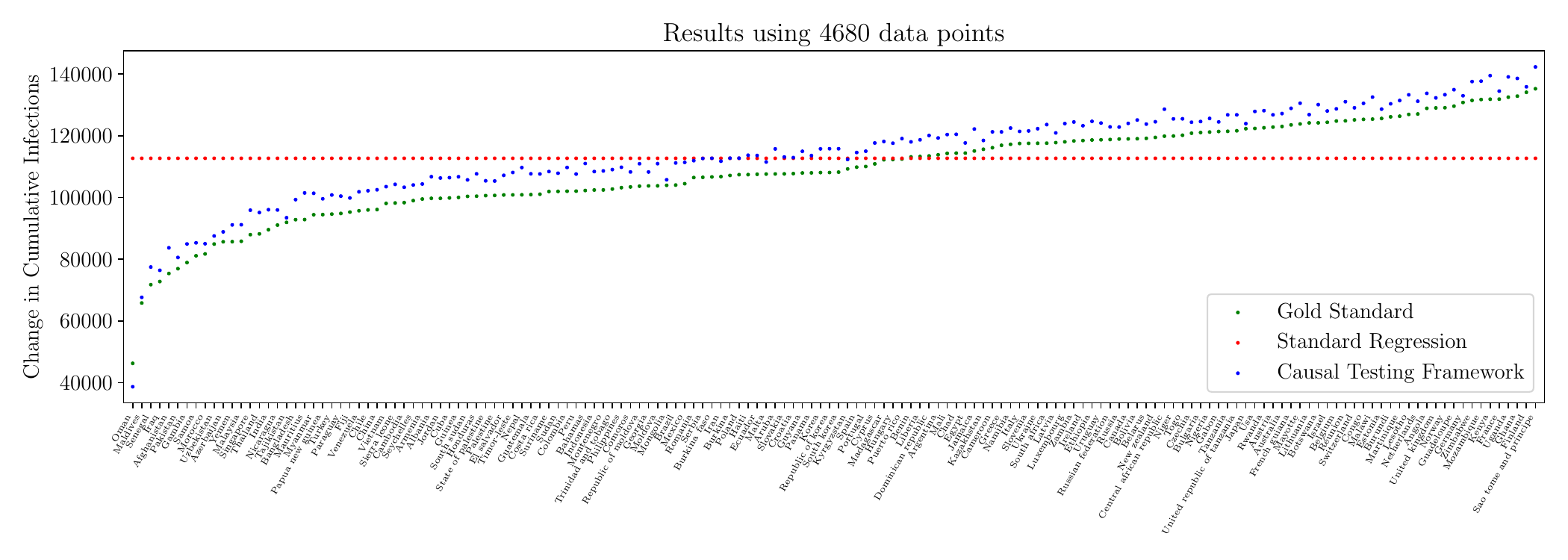}
    \caption{A comparison of the metamorphic test outcomes predicted by the CTF and a naive regression model. The metamorphic test in question increases the value of $\beta$ from 0.016 to 0.02672.}
    \label{fig:covasim-ctf-results}
\end{figure}

\Cref{fig:covasim-ctf-results} summarises the results of applying the CTF to Covasim to predict the effect of increasing transmissibility ($\beta$) on cumulative infections across all locations. These results show three values for each location: (i) the gold standard achieved by applying an SMT approach, (ii) a naive estimate with the simple regression model $I \sim x_0 + x_1 ln(\beta) + x_2 ln(\beta)^2$ (i.e. without employing causal knowledge), and (iii) a causal estimate achieved using the CTF and the approach outlined in this section.

By comparing the CTF results to the gold standard shown in \Cref{fig:covasim-ctf-results}, we can see that the CTF is able to estimate the effect of increasing $\beta$ from $0.016$ to $0.02672$ for each location with reasonable accuracy. Specifically, across the location specific estimates, the CTF has a root mean square percentage error ($RMSPE$) of $0.055$.
This outperforms the naive regression model which provides a uniform prediction that is moderately accurate for `average' locations, but extremely inaccurate for more `extreme' locations ($RMSPE=0.2$).

While these results suggest that the CTF generally overestimates the effect by an average of roughly $5.5\%$ cumulative infections, the overall ordering of the predicted effect sizes is generally consistent with that of the gold standard. We tested this preservation of ordering by calculating the Kendall rank correlation between the (ascending) ordering of the CTF results and the gold standard, returning a value of 0.944 ($p < 0.005$).

\begin{figure}
    \centering
    \includegraphics[width=\linewidth]{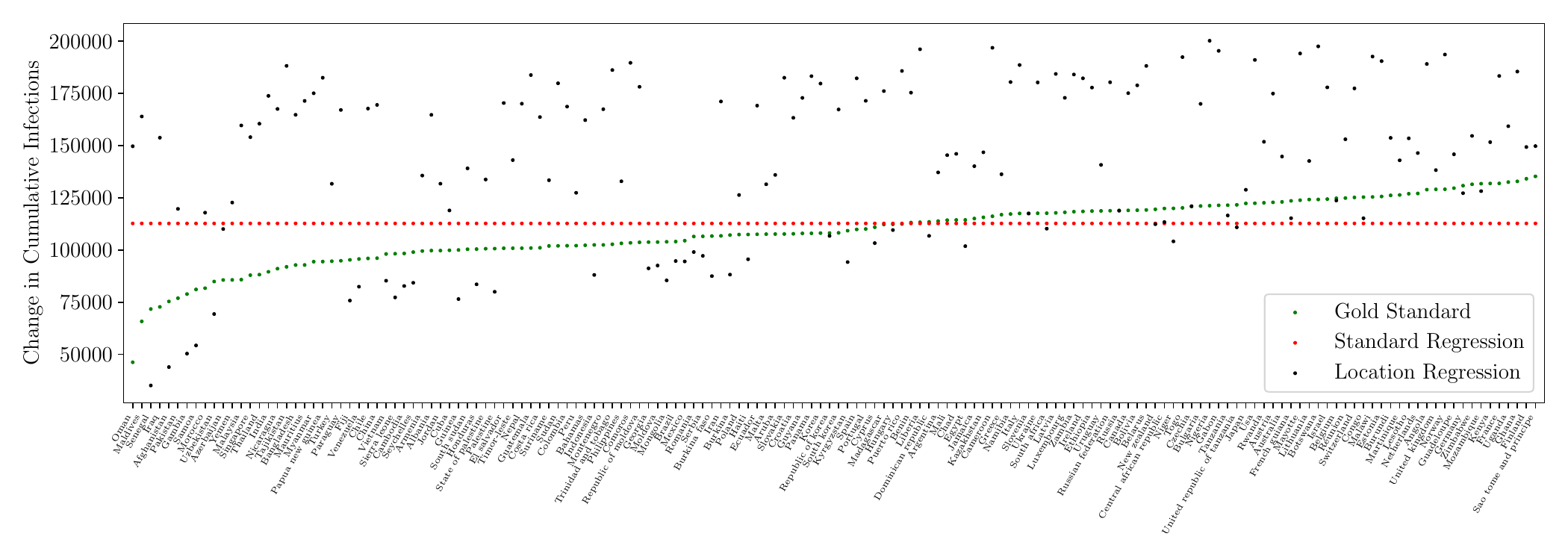}
    \caption{A comparison of the metamorphic test outcomes predicted by a naive regression model and the same model with an interaction between location and $\beta$.}
    \label{fig:covasim-ctf-location-results}
\end{figure}

By contrast, \Cref{fig:covasim-ctf-location-results} shows the results achieved using the smallest adjustment set, $L$, and regression model $I \sim x_0 + x_1 ln(\beta) + x_2 ln(\beta)^2 + ln(x_3) \beta L$. This approach makes location-specific estimates using only the data available for the location in question and is essentially an attempt to apply SMT to incomplete, confounded data. Because each location-specific stratum contains only 30 executions that cover a narrow range of $\beta$ values, the regression model has to make inaccurate extrapolations, leading to significant over- and under-estimates of the true effect ($RMSPE=0.515$) and poor rank preservation, as indicated by a Kendall's rank correlation of $0.228$ ($p<0.005$). This stark contrast in performance highlights the value of employing causal knowledge and domain expertise to use data more efficiently.

\begin{figure}
    \centering
    \includegraphics[width=\linewidth]{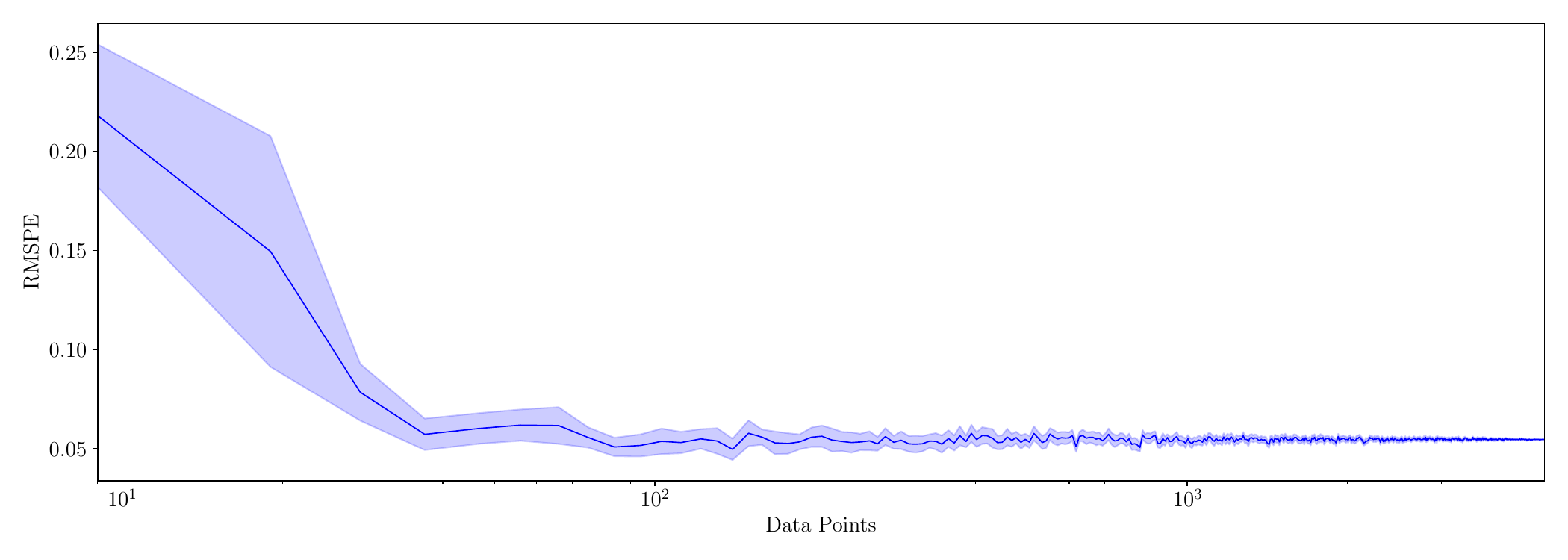}
    \caption{Relationship between root mean square percentage error (RMSPE) of CTF predictions and amount of data used (log scale) with 95\% confidence intervals.}
    \label{fig:covasim-ctf-rmspe-vs-data}
\end{figure}

\begin{figure}
    \centering
    \includegraphics[width=\linewidth]{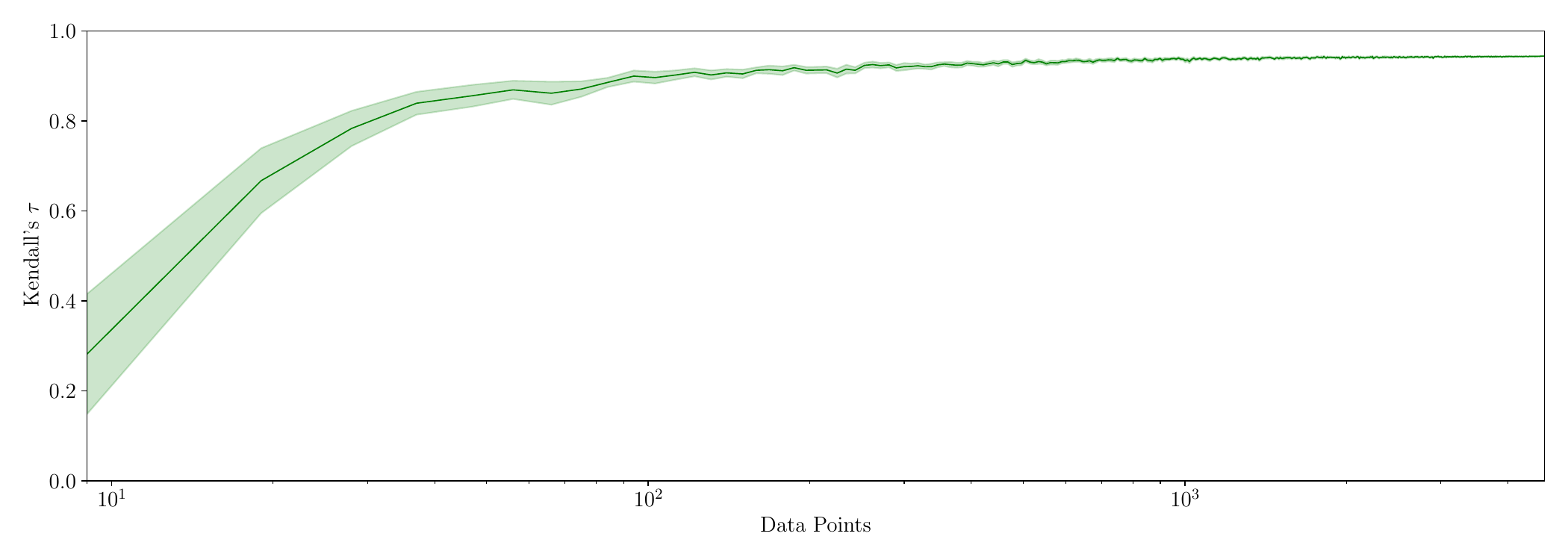}
    \caption{Relationship between Kendall's rank correlation ($\tau$) of CTF predictions and amount of data used (log scale) with 95\% confidence intervals.}
    \label{fig:covasim-tau-vs-data}
\end{figure}

While \Cref{fig:covasim-ctf-results} demonstrates the accuracy with which the CTF can predict SMT outcomes from confounded observational data, these results used the full data set comprising 4680 simulations. Although this is half of the $9360$ executions that would typically be required for a conventional SMT approach, this is still a significant amount of data that may not be available in practice. To investigate how much is necessary in practice, we repeatedly applied the CTF to randomly sampled subsets of the data of decreasing size and calculated the RMSPE and Kendall's rank correlation. We repeated this process 30 times to obtain a distribution of outcomes and report 95\% confidence intervals to demonstrate the error. \Cref{fig:covasim-ctf-rmspe-vs-data} and \Cref{fig:covasim-tau-vs-data} show the results of these experiments. We use a logarithmic scale on the x-axis for these figures as the accuracy changes most significantly between 1 and 200 data points.

\Cref{fig:covasim-ctf-rmspe-vs-data} shows that the RMSPE is greatest with small amounts of data (tens of data points) and quickly reduces to a stable RMSPE of roughly 0.06 by around 200 data points. Similarly, \Cref{fig:covasim-tau-vs-data} shows that the Kendall's rank correlation is initially low (between 0.2 and 0.4) but rapidly increases to a stable value of around 0.9 when 100 to 200 data points are available. This plateau in absolute and comparative error reduction indicates that SMT outcomes can be accurately predicted using only small amounts of data and that larger amounts of data provide negligible gains in accuracy.





This testing scenario has provided evidence for all research questions.

\paragraph{\textbf{RQ1 (Accuracy)}}
\Cref{fig:covasim-ctf-results} shows the accuracy with which the CTF can infer a series of 156 SMT outcomes from confounded observational data \emph{a posteriori}. Although the majority of estimates miss the true effect by around 5.5\%, the ordering of the effect sizes is largely consistent with the gold standard. This finding suggests that, in this case study, the CTF is better suited to drawing comparative conclusions about the effect sizes, such as \emph{``Oman is affected significantly less than Finland''} than absolute conclusions, such as \emph{``Finland observes an increase in cumulative infections of 135829''}.

\paragraph{\textbf{RQ2 (Efficiency)}} 
As shown in \Cref{fig:covasim-ctf-rmspe-vs-data,fig:covasim-tau-vs-data}, after 200 data points, there is negligible improvement to the absolute and comparative accuracy of the estimator. This suggests that, in this case study, the CTF is significantly more efficient than a conventional SMT approach which would require $9360$ executions of the SUT (assuming the source and follow-up tests are repeated 30 times each), with each execution requiring roughly one to two minutes on a moderate specification machine, as noted in earlier in this case study.

\paragraph{\textbf{RQ3 (Practicality)}}
In this case study, we leveraged our limited domain expertise to specify a causal DAG and regression model that facilitates efficient and accurate inference of test outcomes. Most notably, to borrow data from similar locations, we leveraged our knowledge of viral transmission in Covasim to add terms to our regression model for the attributes that influence the effect of transmissibility on cumulative infections, such as contacts and susceptibility. We achieved this using a relatively small DAG containing only eight nodes and employing commonplace regression modelling techniques, such as quadratic, logarithmic, and interaction terms. \\

Overall the findings of this case study highlight the potential offered by a CI-led approach to SMT: whereas a conventional SMT approach would require thousands of carefully controlled executions to test 156 metamorphic relations, the CTF can accurately infer these outcomes from only 200 data points. Furthermore, the CTF enables a tester to infer these outcomes \emph{a posteriori} from potentially confounded data instead of executing the SUT further times. This approach essentially relaxes the constraints ordinarily placed on data used for SMT, facilitating the re-use of existing data while maintaining the ability to draw \emph{causal conclusions}.

\section{Discussion}
\label{sec:discussion}

In this section, we discuss the findings of our three research questions outlined in \Cref{sec:case-studies}, pertaining to the \emph{accuracy}, \emph{efficiency}, and \emph{practicality} of the proposed approach. We also discuss notable additional findings that fall outside the scope of our research questions, including a pair of bugs identified in the case studies.

\subsection{RQ1 (Accuracy): Can we reproduce the results of a conventional MT/SMT approach by applying the CTF to observational data?}

Throughout our case studies, we applied the CTF to a number of different subject systems from different domains to predict MT and SMT outcomes from observational data. That is, data that had not been collected specifically for the testing task in question.

In \Cref{sec:poisson-process-model}, for example, we were able to predict the outcome of two statistical metamorphic tests for a tessellation model with sufficient accuracy to reveal a faulty metamorphic relation. We then confirmed this using a conventional SMT approach. Similarly, in \Cref{sec:cardiac-ap-model}, we predicted several metamorphic test outcomes for a cardiac action potential model, reproducing the results of an existing study. In \Cref{sec:covasim-obs}, we then showed how observational data could be re-used to predict multiple different statistical metamorphic test outcomes for an epidemiological model with high comparative accuracy.

\begin{framed}
  \noindent The CTF is able to accurately reproduce the results of both MT and SMT across a range of scientific modelling software.
\end{framed}

This finding suggests that, by leveraging CI, the CTF can offer an alternative approach to SMT that does not rely on many potentially costly executions of the SUT. Instead, the CTF can be employed \emph{retrospectively} to infer test outcomes from existing, potentially confounded test data, effectively relaxing the constraints ordinarily imposed on the data used for SMT. In this way, the CTF makes it possible to apply SMT where conventional approaches are currently prohibitively expensive, thereby mitigating the problem of long execution times, as discussed in  \Cref{sec:motivating-example} and Kanewala and Bieman's survey \cite{kanewala2014testing}.

While our case studies show that the CTF can infer SMT outcomes with good accuracy for a range of systems, there are more advanced estimation techniques that could be employed to further increase the accuracy. To illustrate this point, in \hyperref[appendix:a]{\textbf{Appendix}} we demonstrate how a more advanced form of regression modelling known as spline regression can more accurately capture the theoretical shape of the cause-effect relationship between $\beta$ (transmissibility) and cumulative infections (originally discussed in \Cref{sec:covasim-obs}). In future work we will compare the performance and usability of more advanced statistical models, such as spline regression \cite{marsh2001spline} and causal forests \cite{athey2019estimating}.

\subsection{RQ2 (Efficiency): In terms of the amount of data required, is the CTF more cost-effective than a conventional MT/SMT approach?}

In \Cref{sec:poisson-process-model} (PLT model) and \Cref{sec:covasim-obs} (Covasim), we used the CTF to conduct SMT using less data than would be required by a conventional SMT approach. In the case of PLT, we were able to reproduce the results of a conventional SMT approach using a fifth of the data, uncovering a failed metamorphic relation in the process. Similarly, in \Cref{sec:covasim-obs} we used the CTF to infer the outcomes of 156 distinct metamorphic relations, as shown in \Cref{fig:covasim-ctf-results}, using roughly half the amount of data required by a conventional SMT approach. We then incrementally reduced the amount of data and repeated our analysis to understand how the accuracy of the approach varies with respect to the amount of data, finding that near-identical results could be achieved using only 200 data points.

Furthermore, although we have not obtained precise timing measurements, we note that the CTF takes roughly a minute to produce all 156 of the location-specific effect estimates shown in \Cref{fig:covasim-ctf-results} on a moderate specification machine. On the other hand, an individual run of Covasim with the settings used in this case study took between one and two minutes on the same machine, and $9360$ executions would be required to test these 156 effects using conventional SMT (with 30 repeats per source and follow-up test case). This would amount to between 156 and 312 hours without parallelisation.

\begin{framed}
  \noindent The CTF is capable of reproducing the results of SMT using significantly less time and data than is required by a conventional SMT approach.
\end{framed}

These findings demonstrate the potential of the CTF to infer the outcomes of metamorphic test cases using significantly less time and data than is required by a conventional SMT approach.
Therefore, in conjunction with our findings for \textbf{RQ1}, our answer to \textbf{RQ2} suggests that the CTF can offer an efficient alternative to conventional MT and SMT approaches that is more compatible with the notoriously demanding properties of scientific software, such as non-deterministic behaviour and long execution times, as described in \Cref{sec:motivating-example}.

An open question surrounding the efficiency of the CTF is how the quality and diversity of the available data affects also the accuracy and scope of inferences. 
To this end, an interesting avenue for future work would be to investigate how existing test generation and selection strategies can be combined with the CTF to generate and prioritise test cases that, once executed, produce execution data with the greatest inferential potential.
In a similar vein, Bareinboim and Pearl \cite{bareinboim2016} have proposed general-purpose methods to combine different data sources generated under different conditions to maximise what can be learned from the data.
Future work could also investigate how these data fusion techniques can be leveraged in a software testing context to further the inferential power of available data sources.

\subsection{RQ3 (Practicality): What practical effort is required from the tester to conduct MT/SMT using the CTF?}
Across our case studies, we primarily drew the causal knowledge necessary to elicit the causal DAGs and regression models from existing studies in which the anticipated cause-effect relationships are well-defined. For example, in \Cref{sec:cardiac-ap-model}, we used the results of an existing study \cite{chang2015bayesian} to specify the causal DAG for the cardiac action potential model (see \Cref{fig:LR91-dag}). Similarly, in \Cref{sec:poisson-process-model} (PLT), we based the shape of our regression models on theoretical results that were also used as the basis of statistical metamorphic relations in the seminal paper on SMT \cite{guderlei2007statistical}. The main expenditure of human effort here was gathering the domain expertise for each system; converting these into causal DAGs was straightforward and required little time. It stands to reason that this would be less time-consuming for a scientific modeller (for example), who would already have a reasonably strong understanding of the underlying subject matter.

As with any model-based testing technique, time and effort are necessary to obtain knowledge and turn it into a domain model. In addition, this process often assumes familiarity with software-specific notions, such as how to characterise a state in a state machine \cite{chow1978testing}, or what events should (or should not) be possible at any given point. Furthermore, the resulting models tend to contain implementation-specific details likely to be unfamiliar to most scientific software users \cite{kanewala2014testing}. By contrast, the CTF relies on an intuitive, domain-agnostic model  (i.e. a causal DAG) that makes essential assumptions transparent and requires a basic understanding of regression modelling. This set of requirements poses a lower barrier to entry for a typical user of scientific software.

More generally, from specification to testing, the components of the CTF outlined in \Cref{sec:causal-testing-framework} assume no prior knowledge of the implementation of the SUT. Instead, the CTF requires the user to specify domain-specific details that are independent of the implementation. This shifts the nature of the burden placed on scientific software testers from being software-specific to domain-specific. In doing so, the CTF facilitates the application of state-of-the-art testing techniques, such as metamorphic testing, to scientific modelling software \emph{without the user even necessarily knowing what a metamorphic relation or test is}. This has been demonstrated throughout the case studies.


\begin{framed}
  \noindent The main expenditure of effort in applying the CTF is the gathering of domain expertise; the task of expressing knowledge in a causal DAG and regression model is comparatively straightforward and involves limited effort. Furthermore, compared to other model-based testing techniques, the barrier to entry for the CTF is better suited to the typical skill set of scientific model users.

\end{framed}

Our work is based on the contention that the effort required to employ the CTF is not unreasonable and that, relative to most model-based testing techniques, the necessary expertise are more familiar to a typical scientific model user \cite{kanewala2014testing}. Namely, the ability to elicit anticipated cause-effect relations in a causal DAG and familiarity with basic regression modelling techniques. However, to precisely quantify and empirically evaluate the feasibility and practicality of the approach, future work will look to conduct a human study in which various scientific developers apply the CTF to a range of scientific software.



\subsection{Summary}
Collectively, our answers to \textbf{RQ1} and \textbf{RQ2} suggest that the CTF offers an accurate and efficient approach that addresses several of the challenges associated with the testing of scientific software outlined by Kanewala and Bieman \cite{kanewala2014testing}. Most notably, through the ability to infer metamorphic test outcomes from small amounts of existing observational data, the CTF mitigates the prohibitively long execution times that typically prevent adequate testing of scientific software. Consequently, the CTF also increases the applicability of metamorphic testing to scientific software, helping to indirectly alleviate the test oracle problem \cite{barr2015}

Of course, the accuracy and efficiency offered by the CTF come at a cost. Our answer to \textbf{RQ3} suggests that the CTF presents a trade-off between practical effort and accuracy/efficiency: by leveraging causal knowledge and domain expertise, the CTF can apply SMT in situations where it is currently impractical. However, these domain expertise can be difficult to obtain for non-domain experts. In the case studies, we found the main expenditure of human effort to be in collecting the domain expertise necessary to apply the techniques; the process of converting these into a DAG and regression model required considerably less effort.

\subsection{Additional Findings}
Throughout our case studies, we also identified a number of additional findings that warrant discussion. First, we discuss the need for explainability and how causal DAGs help to address this. Second, we discuss a pair of bugs identified in the case studies using the CTF.

\paragraph{Explainability}
When testing scientific software, the reasoning behind a particular test passing or failing (i.e. the test oracle procedure) is rarely made explicit. For example, modellers often use regression testing to check whether changes to the SUT have affected model predictions or results. Any deviations are then typically validated by a domain expert. This form of ad-hoc validation lacks transparency and, as such, cannot be easily interrogated by prospective users of the SUT. For applications such as infectious disease modelling, where software outputs may inform important policy decisions, there is a need for accountable and explainable test results. Explainability is also a topic of growing concern in fields such as healthcare \cite{holzinger2019causability} that are increasingly using black-box machine learning techniques but require transparent, accountable, and interpretable decision making \cite{burkart2021survey}.

To this end, the CTF incorporates \emph{explainability} into the testing process. Specifically, by utilising causal DAGs for CI, the CTF includes a lightweight and transparent artefact that partially explains the reasoning behind reaching a particular test outcome (i.e. why a specific adjustment set, and therefore statistical model, yields a \emph{causal} estimate). Furthermore, the causal test case (\Cref{def:causal-test-case}) includes an explicit test oracle (\Cref{def:causal-test-oracle}) that captures `correctness' in terms of some causal metric, such as the $ATE$ or $RR$. Both assets can be easily accessed and interrogated, increasing the explainability and reputability of tests.

With this built-in notion of explainability, we posit that the CTF also has the potential to complement existing techniques in the scientific modelling context that often rely on implicit domain expertise for testing, such as regression testing. However, the causal DAG and test oracle do not communicate all assumptions with the potential to influence test results and their interpretation. For example, the anticipated functional form of a particular cause-effect relationship will influence the design of the regression model and its resulting predictions. A potential avenue for future work would be to investigate methods for improving the explainability of the CTF. For example, one could look into more expressive graphical models of causality that capture the expected functional form.

\paragraph{Bugs Found}
Our case studies also revealed two interesting, previously undiscovered bugs in two of the studied scientific models: the Poisson Line Tessellation model and Covasim.

First, in \Cref{sec:poisson-process-model}, we found that the relationship between intensity and number of polygons per unit area described in \cite{guderlei2007statistical} was more fragile at smaller window sizes. This suggested that the window size (width and/or height) has a causal effect on the number of polygons per unit area, while \cite{guderlei2007statistical} stated that these variables should be independent. We then designed a causal test case to confirm that increasing the window width from 1 to 2 whilst holding intensity constant \emph{caused} a significant change in the number of polygons per unit area.

Second, in \Cref{sec:covasim-exp}, we found a bug in Covasim's vaccine implementation where, upon prioritising the elderly for vaccination, the number of vaccinated individuals grew to nearly ten times the number of individuals in the simulation. While this does not appear to significantly affect the key outputs of the model, it is not difficult to imagine how such a bug could lead to an overestimation of the effects of interventions.

\subsection{Threats to Validity}
Our evaluative case studies in \Cref{sec:case-studies} do not claim to make generalisable conclusions regarding the accuracy, efficiency, and effort associated with the CTF. Instead, these case studies serve as proofs of concept that show - for the studied subject systems - how formulating metamorphic testing as a CI problem makes it possible to apply the approach in situations where conventional metamorphic testing methods are impractical. Nonetheless, there are some threats to validity worth considering here.

\subsubsection{External Validity}
In this work, the main threat to external validity is that our case studies only cover three subject systems involving a moderate number of input and output variables. As graphical CI requires domain expertise for the data-generating mechanism in the form of a causal DAG, a significant amount of time was spent familiarising ourselves with the subject systems and understanding their constituent cause-effect relationships. As a result, this limited our ability to systematically collect and analyse large numbers of varied subject systems.

Furthermore, our subject systems were all implemented in Python. Therefore, our findings do not necessarily generalise to scientific modelling software implemented in other languages. However, the CTF only requires execution data in CSV format to perform causal testing observationally and can thus be applied, in theory, to tabular data produced by \emph{any} scientific model.

As a consequence of the aforementioned threats to external validity, we acknowledge that our results may not generalise to \emph{all} forms of scientific modelling software. However, we attempt to mitigate the aforementioned threats to external validity by selecting models that differ in their complexity, subject matter, and modelling paradigm. In addition, as discussed in \Cref{sec:case-studies}, the selected systems have important but vastly different applications across a variety of domains, and have all been the subject of prior research.

\subsubsection{Internal Validity}
In this paper, the main threat to internal validity is that we did not optimise the estimators and configuration parameters thereof for our case studies. While this avoids the problem of over-fitting, it means there may exist statistical models that are more suitable for modelling and inferring the behaviour of the input-output relationships under study.

In the same vein, we specified regression equations that capture the expected functional form of various input-output relationships. For example, when testing Covasim in \Cref{sec:covasim-obs}, we specified a regression model which captures our broad understanding of how cumulative infections vary with various causally relevant parameters. We called upon our experience with the models and subject area to specify these equations. However, different domain experts may have different opinions about the correct functional forms of the input-out relationships and may therefore have specified these relationships differently or more accurately.

As a consequence of the above threats to internal validity, we acknowledge that there alternative statistical models may achieve more precise causal inferences for the subject systems. However, we partially mitigate the above threats to internal validity by manually inspecting the functional forms of the relationships between inputs and outputs of interest in the SUT. We achieve this by varying one parameter at a time and observing how the output in question changes in response (in a similar way to our sensitivity analysis case study in \Cref{sec:cardiac-ap-model}). We also include a more advanced regression model in \hyperref[appendix:a]{\textbf{Appendix}} that more accurately captures the relationship between transmissibility ($\beta$) and the number of cumulative infections in Covasim.







\section{Related Work}
\label{sec:related-work}
In this section, we provide a brief review of work related to the two main topics concerning our paper: approaches for testing scientific software and causality in software testing. Additionally, we summarise automatic approaches to generating causal DAGs and highlight a number of open research challenges.

\subsection{Testing Techniques for Scientific Software}
As stated in Kanewala and Bieman's survey \cite{kanewala2014testing}, scientific models are seldom tested using systematic approaches. Instead, techniques such as sensitivity \cite{oakley2004probabilistic} and uncertainty analysis \cite{farajpour2013error} are often employed to analyse and appraise scientific models. However, these approaches generally require many costly executions that make them prohibitively expensive at scale \cite{conti2010bayesian}. To address this issue, modellers have turned to emulator approaches \cite{rasmussen2003gaussian, conti2010bayesian}, where a surrogate model is developed to approximate the behaviour of the simulation and provide an efficient way to validate behaviour \cite{chang2015bayesian, vernon2014galaxy}. However, these emulators are driven by statistical associations and are unable to draw \emph{causal} inferences from existing test data.

Another issue that precludes the testing of scientific modelling software is the oracle problem \cite{barr2015}; the lack of a mechanism that can be used to ascertain whether the outcome of a test case is correct or not. Kanewala and Bieman's survey \cite{kanewala2014testing} identifies several approaches followed by scientific modellers to overcome the oracle problem, including: pseudo oracles, comparison to analytical solutions or experimental results, and expert judgement. In addition to these solutions, modellers have also turned to metamorphic testing (see \Cref{sec:background}) to overcome the lack of oracle. This approach relies on the scientists being able to specify metamorphic relations capable of revealing faults. However, these relationships are notoriously challenging to identify \cite{segura2016survey}.

To assist with the identification of metamorphic relations, Kanewala and Bieman developed a machine learning approach for predicting metamorphic relations for numerical software \cite{kanewala2013using}. This is achieved by representing numerical functions as a statement-level control flow graph and extracting features from this graph to train a classifier. In recent years, several new approaches for automatically predicting metamorphic relations for a specific form of software have been proposed, including for cyber-physical systems \cite{ayerdi2021generating, ayerdi2022evolutionary} and matrix calculation programs \cite{rahman2018predicting}. However, the generation of metamorphic relations remains a difficult problem with automatic solutions available for only a few specific forms of software.

\subsection{Causality in Software Testing}
In more conventional settings, CI techniques have been applied to the software testing problem of fault localisation (FL). Informally, FL concerns identifying locations of faults in a program \cite{wong2016survey} and often involves computing a ``suspiciousness metric'' for software components, such as program statements. However, these metrics are often confounded by other software components. To address this, Baah et al. \cite{baah2010causal} translated FL to a CI problem, using program dependence graphs as a model of causality to estimate the causal effects of program statements on the occurrence of faults. Subsequent papers build on this to handle additional sources of bias \cite{baah2011mitigating}; leverage more advanced statistical models \cite{baah2011mitigating, podgurski2020counterfault}; and adapt to different software components \cite{shu2013mfl, bai2015numfl, gore2012reducing, podgurski2020counterfault}.

More recently, Lee et al. have introduced the Causal Program Dependence Analysis Framework and applied it to FL. This is a CI-driven framework that measures the strength of dependence between program elements by modelling their causal structure \cite{lee2021causal}. Unlike previous CI-based FL techniques, this framework does not use static analysis to construct its underlying model of causality, and instead approximates the causal structure by observing the effects of interventions. In a series of experiments, the framework has been shown to outperform slicing-based and search-based FL techniques, and help developers focus on key dependencies. Furthermore, due to its focus on dependence relations instead of coverage, it is less susceptible to coincidental correctness (executions that pass but cover faulty components).

In a similar vein, software testing often involves understanding \emph{why} a particular outcome occurs, such as a program failure. To this end, Johnson et al. \cite{johnson2020causal}, developed a tool that explains the root cause of faulty software behaviour. This tool creates ``causal experiments'' by mutating an existing test to form a suite of minimally different tests that, contrary to the original, are not fault-causing. The passing and failing tests can then be compared to understand \emph{why} a fault occurred. Similarly, Chockler et al. \cite{chockler2021compositional} developed a tool to \emph{explain} the decisions of deep neural network (DNN) image classifiers. Following the actual causes framework \cite{halpern2005causes}, this tool offers explanations in the form of minimal subsets of pixels sufficient for the DNN to classify an image.

Another software testing technique concerning causality is cause-effect graphing, a black-box approach adapted from hardware testing. Here, input-output relationships are expressed in a variant of a combinatorial logic network, known as a cause-effect graph, created by manually extracting causes, effects, and boolean constraints from natural language specifications \cite{nursimulu1995cause, myers2004art}. Unlike the previous techniques, this approach does not use CI.

Recent work presented in \cite{giamattei2023} frames software testing in terms of causal reasoning. The authors conceptualise an iterative approach for test case generation in which test cases and the causal DAG are generated together and used to improve each other. However, the work is still at a preliminary stage, and the important link between CI and metamorphic testing is not discussed.

\subsection{Automatic Generation of Causal DAGs}
\label{sub:generatedDags}
In this paper, we have assumed that all causal DAGs are specified manually by a domain expert. While this is an intuitive and widely accepted approach for conducting CI in fields such as epidemiology and social sciences, there are two potential methods that could, in theory, (partially) automate this process.

First, under certain strict assumptions and with large quantities of data, it is possible to predict the structure of causal DAGs from observational data. Where \emph{model inference} provides a source of models for traditional MBT techniques \cite{UttingPL12}, the field of \emph{causal discovery} (CD) \cite{malinsky2018causal} provides methods to infer causal structures from data by exploiting asymmetries that distinguish association from causation \cite{glymour2019review}. However, due to the need for large amounts of data and their strict assumptions, we have had limited success in applying CD algorithms to model execution data. We plan to investigate this route further in future work.

Second, causal DAGs can be generated via static analysis of source code. DAGs derived in this way have already been used for FL \cite{podgurski2020counterfault, lee2021causal}. However, this approach relies on source code being openly available and produces a detailed, low-level model of causality for the SUT. While this level of granularity is ideal for the purpose of FL, the resulting causal DAG would be less suitable for a typical scientific modeller.

In addition to the aforementioned challenges, there is a fundamental barrier to using automatically generated models of causality for testing: inferred models represent the implemented system rather than the true specification. In other words, even if we could perfectly recover the DAG of the implementation, this would contain any bugs the implementation may have. We would, in effect, be testing the system against itself, so it would trivially look correct. Hence, the correctness of any inferred DAGs must be verified by a domain expert.

\subsection{Machine Learning-Inferred Models of Tested Behaviour}
In this work, we employ causality-informed linear regression models to infer metamorphic test outcomes. This aspect of our work relates to a significant body of work on machine learning approaches for inferring models from test executions. While Weyuker started this line of research 40 years ago \cite{weyuker1983assessing}, it has become particularly active in the last decade.

Most testing approaches that incorporate machine learning do so in the context of regression testing, where the inferred model represents the correct behaviour that can be used to identify any faults arising in subsequent software versions. Such approaches often use off-the-shelf machine learning and regression algorithms, chosen to fit the characteristics of the software behaviour in question. These have included standard linear regression \cite{arrieta2021using}, state machine inference \cite{walkinshaw2016inferring}, and decision trees \cite{briand2009using} amongst others. 

Such approaches are applicable to situations where (a) there is an established, reasonably correct system in place to derive tests from, and (b) there is a sufficiently large and diverse amount of execution data available. In our case, neither of these conditions holds. The computational models we analyse are exploratory in nature, and would not serve as a reliable oracle in their own right. Instead, we depend on causal properties provided by the developer in the DAG. Furthermore, computational models are subject to the various restrictions described in Section \ref{sec:motivating-example} - namely, high execution times, large and complex input spaces, and high computational costs. These restrictions prevent us from collecting a set of executions that is sufficiently large and diverse to accurately characterise the underlying behaviour.

\section{Conclusion and Future Work}\label{sec:conclusion}
In this paper, we presented the Causal Testing Framework (CTF): a conceptual framework that facilitates the application of causal inference (CI) techniques to software testing problems.
This framework follows a model-based testing approach to incorporate an explicit model of causality into the software testing process in the form of a causal DAG, enabling the direct application of graphical CI methods to software testing activities.
Due to its fundamentally causal nature, we took a particular focus on metamorphic testing in this work.

A key contribution of the CTF is its ability to infer metamorphic test outcomes from previous execution data, despite the presence of confounding, providing an efficient method for testing scientific models in situations where it is currently impractical or infeasible.
To demonstrate this, we applied our open source reference implementation of the CTF to three real-world scientific models of varying size and complexity, including a Poisson line tessellation model, a cardiac action potential model, and an epidemiological agent-based model.
The results of these case studies suggest that, through the use of CI, the CTF can accurately infer metamorphic test outcomes from existing test data using significantly less data than is required by a conventional statistical metamorphic testing approach.


Software testing is an inherently causal process, and the field of CI holds much-untapped potential. To this end, the CTF lays the foundation for a new line of causality-driven software testing techniques. 
In one line of future work, we plan to apply the CTF to more causality-led testing activities, such as regression testing and A/B testing, to better understand how CI can support different testing activities. 
A separate direction of research would be to establish a (semi-)automatic, reliable process for the discovery of causal DAGs representing software systems. Such an artefact could be used as a starting point for a causal specification, reducing the amount of human effort required to apply the CTF and thus lower the barrier to entry.  
\section*{Acknowledgements}
Foster, Walkinshaw, and Hierons are funded by the EPSRC CITCoM grant EP/T030526/1.
For the purpose of open access, the author has applied a Creative Commons Attribution (CC BY)\footnote{Where permitted by UKRI a CC-BY-ND licence may be stated instead.} licence to any Author Accepted Manuscript version arising.
\bibliographystyle{ieeetr}
\bibliography{bibliography}
\appendix
\section*{Appendix}
\label{appendix:a}

\subsection*{A more advanced regression model for Covasim}
In \Cref{sec:covasim-obs}, we designed a regression model that broadly captures the expected relationship between cumulative infections and various causally relevant parameters, such as transmissibility $\beta$ and household contacts $C_H$. This regression model uses conventional regression modelling techniques to specify the relationships of interest. Namely, quadratic terms, log transformations, and effect modifiers.

However, this model does not capture the relationship between $\beta$ and cumulative infections perfectly because the relationship follows a sigmoid function (i.e. a characteristic S-shaped curve). Informally, we can explain this relationship as follows. Initially, when $\beta$ is low, there are few infections because the rate of viral transmission is low. Then, as $\beta$ increases past some critical threshold, an exponential growth in the transmission rate occurs. Eventually, enough of the population becomes infected and gains immunity or dies, rapidly reducing the rate of viral transmission. This sudden reduction causes cumulative infections to level off, completing the characteristic S shape.

One of the weaknesses of polynomial regression is its unpredictable tail behaviour \cite{wolf2013sage}. This limitation is particularly problematic for modelling sigmoid relationships, where the tails are necessarily flat. To address this limitation, we employed a more advanced form of regression known as spline regression \cite{marsh2001spline}.

In short, spline regression involves constructing a piece-wise polynomial over contiguous regions of the data. Within each region, a separate polynomial function of degree $n$ is fit to the subset of data. This approach to regression essentially breaks the problem into discrete stages and is an effective technique for capturing non-linear relationships. In many cases, a third-degree polynomial is used to model each region, in which case the resulting splines are referred to as cubic splines.

Based on our limited domain expertise, to capture the sigmoid relationship between $\beta$ and cumulative infections, we used cubic splines with two (internal) knots. With this approach, our aim was to separate the data into three regions corresponding to the three distinct phases of the sigmoid function described above (initial slow growth in infections, exponential growth, and plateau in infections). 

\begin{figure}[!h]
    \centering
    \includegraphics[width=\linewidth]{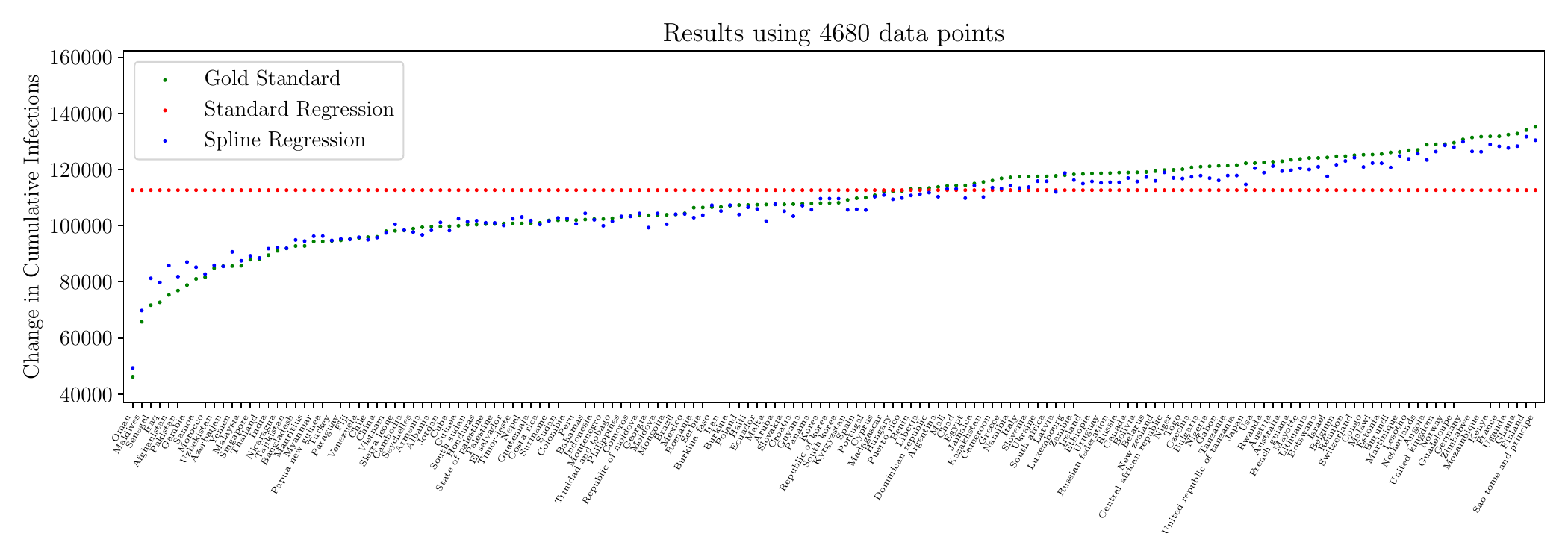}
    \caption{A comparison of the metamorphic test outcomes predicted by a cubic spline regression with two knots and the naive regression.}
    \label{fig:covasim-spline-results}
\end{figure}

\Cref{fig:covasim-spline-results} shows the metamorphic test outcomes predicted using cubic spline regression. From an informal visual inspection, it is clear that the majority of estimates are more accurate than the previous regression model, which generally overestimated the effects and had a root mean square percentage error ($RMSPE$ of 0.055) and a Kendall's rank correlation of $0.944$ ($p<0.005$). By contrast, the cubic spline approach had an $RMPSE$ of 0.032 and a Kendall's rank correlation of $0.915$ ($p<0.005$). Therefore, the spline regression technique provided better absolute accuracy (indicated by $RMSPE$), but worse comparative accuracy (indicated by Kendall's rank correlation). The performance of both approaches could likely be improved by a domain expert who may have a more precise characterisation of the anticipated relationships.

We decided not to include the cubic splines approach in the case studies, as it requires more advanced statistical modelling knowledge that is unlikely to be commonplace to prospective users. However, it is worth including as an appendix because it introduces a potentially valuable trade-off. Namely, more advanced, semi-parametric statistical estimators can be employed with arguably less domain knowledge to learn intricate shapes from the available data. However, this introduces an additional burden: the need for expertise in such modelling techniques. 

Overall, in this example, we were able to configure the spline regression model in a logical way that is justified by domain expertise (i.e. splitting the relationship into three key regions, each of which can be modelled with a cubic polynomial). This shows how more advanced statistical means can be employed to achieve better results. In future work, we will investigate the application of other semi- and non-parametric statistical models within the Causal Testing Framework.
\end{document}